\documentclass[a4paper,fleqn,usenatbib]{mnras}
\usepackage{mathptmx}

\usepackage[T1]{fontenc}
\usepackage{ae,aecompl}

\usepackage[latin1]{inputenc}
\usepackage{multirow}
\usepackage{graphicx} 
\usepackage{multicol}
\usepackage{float}
\usepackage{footnote}
\usepackage{amssymb} 	
\usepackage[figuresright]{rotating}
\usepackage{amsmath} 	
\usepackage{url}
\usepackage{rotating}
\usepackage{caption}

\usepackage{subcaption} 
\usepackage[export]{adjustbox}
\usepackage[dvipsnames]{xcolor} 

\title[High-resolution radio observations of Perseus]{High-resolution VLA low radio frequency observations of the Perseus cluster: radio lobes, mini-halo and bent-jet radio galaxies}
\author[M. Gendron-Marsolais et al.]{M. Gendron-Marsolais$^{1}$\thanks{E-mail: mgendron@eso.org}, 
J. Hlavacek-Larrondo$^{2}$,
R. J. van Weeren$^{3}$, 
L. Rudnick$^{4}$,
\newauthor T. E. Clarke$^{5}$, 
B. Sebastian$^{6}$,
T. Mroczkowski$^{7}$,
A. C. Fabian$^{8}$, 
K. M. Blundell$^{9}$, 
\newauthor  E. Sheldahl$^{10}$,
K. Nyland$^{11}$,
J. S. Sanders$^{12}$,
W. M. Peters$^{5}$,
and H. T. Intema$^{3}$\\
$^{1}$European Southern Observatory, Alonso de Co\'ordova 3107, Vitacura, Casilla 19001, Santiago de Chile  \\
$^{2}$D\'epartement de Physique, Universit\'e de Montr\'eal, Montr\'eal, QC H3C 3J7, Canada\\
$^{3}$Leiden Observatory, Leiden University, Niels Bohrweg 2, NL-2333CA, Leiden, The Netherlands\\
$^{4}$Minnesota Institute for Astrophysics, School of Physics and Astronomy, University of Minnesota, 116 Church Street SE, Minneapolis, \\ MN 55455, USA \\
$^{5}$Naval Research Laboratory, Code 7213, 4555 Overlook Ave. SW, Washington, DC 20375, USA\\
$^{6}$National Center for Radio Astrophysics - Tata Institute of Fundamental Research Post Box 3, Ganeshkhind P.O., Pune 41007, India \\
$^{7}$European Southern Observatory, Karl-Schwarzschild-Str. 2, DE-85748 Garching b. M\"unchen, Germany \\
$^{8}$Institute of Astronomy, University of Cambridge, Madingley Road, Cambridge CB3 0HA \\
$^{9}$University of Oxford, Astrophysics, Keble Road, Oxford OX1 3RH, UK\\
$^{10}$University of New Mexico, Albuquerque, NM, USA\\
$^{11}$ National Research Council, resident at the U.S. Naval Research Laboratory, 4555 Overlook Ave. SW, Washington, DC 20375, USA\\
$^{12}$Max-Planck-Institut f\"ur extraterrestrische Physik, 85748 Garching, Germany 
}

\date{Accepted for publication in MNRAS}

\pubyear{2020}

\begin{document}
\label{firstpage}
\pagerange{\pageref{firstpage}--\pageref{lastpage}}
\maketitle



\begin{abstract}
We present the first high-resolution 230-470 MHz map of the Perseus cluster obtained with the Karl G. Jansky Very Large Array. The high dynamic range and resolution achieved has allowed the identification of previously-unknown structures in this nearby galaxy cluster. New hints of sub-structures appear in the inner radio lobes of the brightest cluster galaxy NGC 1275. The spurs of radio emission extending into the outer X-ray cavities, inflated by past nuclear outbursts, are seen for the first time at these frequencies, consistent with spectral aging. Beyond NGC 1275, we also analyze complex radio sources harbored in the cluster. Two new distinct, narrowly-collimated jets are visible in IC 310, consistent with a highly-projected narrow-angle tail radio galaxy infalling into the cluster. We show how this is in agreement with its blazar-like behavior, implying that blazars and bent-jet radio galaxies are not mutually exclusive. We report the presence of filamentary structures across the entire tail of NGC 1265, including two new pairs of long filaments in the faintest bent extension of the tail. Such filaments have been seen in other cluster radio sources such as relics and radio lobes, indicating that there may be a fundamental connection between all these radio structures. We resolve the very narrow and straight tail of CR 15 without indication of double jets, so that the interpretation of such head-tail sources is yet unclear. Finally, we note that only the brightest western parts of the mini-halo remain, near NGC 1272 and its bent double jets.
\end{abstract}

\begin{keywords}
Galaxies: clusters: individual: Perseus cluster - galaxies: jets - radio continuum: galaxies - X-rays: galaxies: clusters - cooling flows 
\end{keywords}

\section{Introduction}

Cluster environments host numerous remarkable phenomena as they form the largest gravitationally bound structures in our universe.
The X-ray emitting intracluster medium (ICM) can be perturbed both internally by the outburst of the active galactic nuclei (AGN) of the dominant galaxy and externally from interactions with other clusters, groups, and individual galaxies
The Perseus cluster is a classic example of such an environment. Being the brightest cluster in the X-ray sky, it is a well-studied object and has been observed extensively across all the electromagnetic spectrum.

Deep  X-ray observations of the Perseus cluster have revealed several types of disturbances in the ICM.
At least two pairs of X-ray cavities have been identified: the first pair, located at $5 < r <20$ kpc from the AGN, are filled with radio emission \citep{bohringer_rosat_1993} and the second, located further out at $25 < r < 45$ kpc from the AGN, are devoid of radio emission above $\gtrsim1$ GHz \citep{branduardi-raymont_soft_1981,fabian_distribution_1981,churazov_asymmetric_2000}. 
Inflated by generations of outbursts from the central AGN, these constitute the imprint of radio mode feedback, injecting energy into the ICM and compensating its radiative losses. 
Beyond these cavities are ripple-like quasi-spherical structures interpreted as sound waves which may be responsible for part of the transport and dispersion of this energy throughout the cooling region \citep{fabian_deep_2003}.
Shocks surrounding the inner cavities have also been identified in \cite{fabian_very_2006}.
A large semicircular cold front is seen $\sim100$ kpc west of the nucleus \citep{fabian_wide_2011}, and another one is also detected much further out, $\sim730$ kpc east of the nucleus \citep{simionescu_large-scale_2012,walker_split_2018}. 
These structures are interpreted as the result of perturbations of the cluster's gravitational potential well from a minor merger.
Similar to cold fronts but with the opposite curvature, a large region of weak X-ray emission located $\sim100$ kpc south of the nucleus has been identified as a ``bay'' \citep{fabian_wide_2011}. 
It could be the imprint of a giant Kelvin-Helmholtz instability \citep{walker_is_2017}.

Besides the radio emission directly originating from the central AGN and its jets, a faint diffuse radio component is also present at lower frequencies ($\lesssim 1$GHz, \citealt{soboleva_3c84_1983,pedlar_radio_1990,burns_where_1992,sijbring_radio_1993}), classified as a ``mini-halo''.
Mini-halos are faint and diffuse radio structures with steep spectra, filling the cooling core of some relaxed clusters (e.g. \citealt{feretti_clusters_2012,giacintucci_occurrence_2017}). Their origin remains unclear since the radiative timescale of the relativistic electrons coming from the central AGN  is much shorter than the time required for them to reach the extent of the mini-halo. Two possible mechanisms are debated to explain the mini-halo emission: the hadronic or the turbulent re-acceleration models. According to these, mini-halos originate either from the generation of new particles from inelastic collisions between relativistic cosmic-ray protons and thermal protons (e.g. \citealt{pfrommer_constraining_2004}), or from the reacceleration of pre-existing electrons by turbulence \citep{gitti_modeling_2002,gitti_particle_2004}.
About thirty mini-halos are known so far and most have irregular morphologies extending on $\sim 100$ kpc scales (e.g. \citealt{giacintucci_new_2014,giacintucci_occurrence_2017,giacintucci_expanding_2019,richard-laferriere_relation_2020}). 
Recent deep \textit{NSF's Karl G. Jansky Very Large Array} (VLA) observations of the Perseus cluster at 230-470 MHz have however revealed a wealth of inner structures in the mini-halo \citep{gendron-marsolais_deep_2017}. Mostly confined behind the western cold front, it also shows several radial filaments, a concave radio structure associated with the southern X-ray bay and sharp radio edges that correlate with X-ray edges.

The Perseus cluster also hosts several other radio galaxies with complex morphologies, including NGC 1265, NGC 1272, CR 15 and potentially IC 310 (e.g. \citealt{sijbring_multifrequency_1998}). Due to their motion through the cluster, these galaxies have bent radio jets caused by the ram pressure from the ICM (e.g. \citealt{jones_hot_1979,begelman_twin-jet_1979}).

These multi-wavelength studies of the Perseus cluster have fundamentally changed our understanding of cluster's environment. Among these, the results presented in \cite{gendron-marsolais_deep_2017} showed the capability of VLA data at 230-470 MHz to provide a detailed view of the radio structures in Perseus by revealing a multitude of new structures associated with the mini-halo. We now extend this study to higher spatial resolutions and investigate the smaller scales of the various radio structures at the same low radio frequencies. In this article, we present new, deep VLA high-resolution observations of the Perseus cluster at 230-470 MHz. 
These A-configuration observations reach a resolution more than 4 times higher than the B-configuration observations at the same frequencies presented in \cite{gendron-marsolais_deep_2017} and constitute the highest resolution observations of the Perseus cluster below 1 GHz. 

In Section \ref{VLA Observations and data reduction_highres}, we present the observations used in this work and the data reduction of the VLA dataset. Sections \ref{Results_highres}, \ref{NGC 1275_highres}, \ref{Head-tail sources in the Perseus cluster_highres} and \ref{Spectral index maps of NGC 1265 and IC 310} discuss the different structures found in the radio observations. Results are summarized in Section \ref{Conclusion_highres}.

We assume a redshift of $z = 0.0183$ for NGC 1275, the brightest cluster galaxy (BCG, \citealt{forman_observations_1972}), corresponding to a luminosity distance of 78.4 Mpc, assuming $H_{0} = 69.6 \text{ km s}^{-1} \text{Mpc}^{-1}$, $\Omega_{\rm M} = 0.286$ and $\Omega_{\rm vac} = 0.714$. This corresponds to an angular scale of 22.5 kpc arcmin$^{-1}$.

\section{Observations and data reduction}\label{VLA Observations and data reduction_highres}

\subsection{$P$-band (230-470 MHz) VLA observations}

The Perseus cluster has been extensively imaged with the pre-upgrade VLA at all frequencies and configurations. Here we present the first $P$-band (230-470 MHz) observations of Perseus taken with the upgraded VLA (following the \textit{Expanded Very Large Array Project}, \citealt{perley_expanded_2011}) in A-configuration (project $13B-026$, PI Hlavacek-Larrondo). The data were taken on May 16, 2014 for a total of 5 hours. The A-configuration is the most extended of the VLA, yielding to the highest resolution, with typical synthesized beam sizes of $\sim 5 \arcsec$ and maximum recoverable scales of $\sim 3 \arcmin$ at $P$-band frequencies. 
We also obtained B-configuration observations (synthesized beam sizes of $\sim 15 \arcsec$ and largest recoverable scales of $\sim 10 \arcmin$) that have been the subject of a recent publication (see \citealt{gendron-marsolais_deep_2017}), focusing on the mini-halo emission.

The A-configuration observations were taken with 27 operational antennas.  
Antenna 14 was flagged at the beginning of the data reduction process as problems with the $P$-band receiver during the observation period were reported in the operator log.
The $P$-band receivers have 16 spectral windows, each comprising 128 channels with a width of 125 kHz, covering 230 to 470 MHz. The dataset consists of a total of 58 scans, starting with 10 mins on 3C48 (for flux, delay and bandpass calibrations), 10 mins on 3C147 in the middle of the observation period and a final 10 mins on 3C147 at the end (both for delay and bandpass calibrations). The rest of the observations consist of scans of about 5 mins each centered on NGC 1275.

The data reduction was performed with the package Common Astronomy Software Applications (CASA, version 4.7, \citealt{mcmullin_casa_2007}). We have developed a pipeline to account for the strong presence of radio frequency interference (RFI) at these low-frequencies and the extremely bright central AGN in Perseus. The steps of the data reduction process have been detailed in \cite{gendron-marsolais_deep_2017}. Here we have applied the same pipeline to the A-configuration observations.
In summary, the data calibration was performed separately on each spectral window.
The task \textsc{fixlowband} was applied to correct the labelling of the $P$-band system polarizations. 
The most apparent RFI were removed using the mode \textsc{tfcrop} from the task \textsc{flagdata} and the flagger framework \textsc{AOFlagger} \citep{offringa_morphological_2012}.
Both 3C48 and 3C147 were used as bandpass calibrators to increase the signal-to-noise ratio on the bandpass solutions.
Outlier solutions in calibration tables were removed manually.

\begin{center}
\begin{figure*}
\centering
\includegraphics[width=\textwidth]{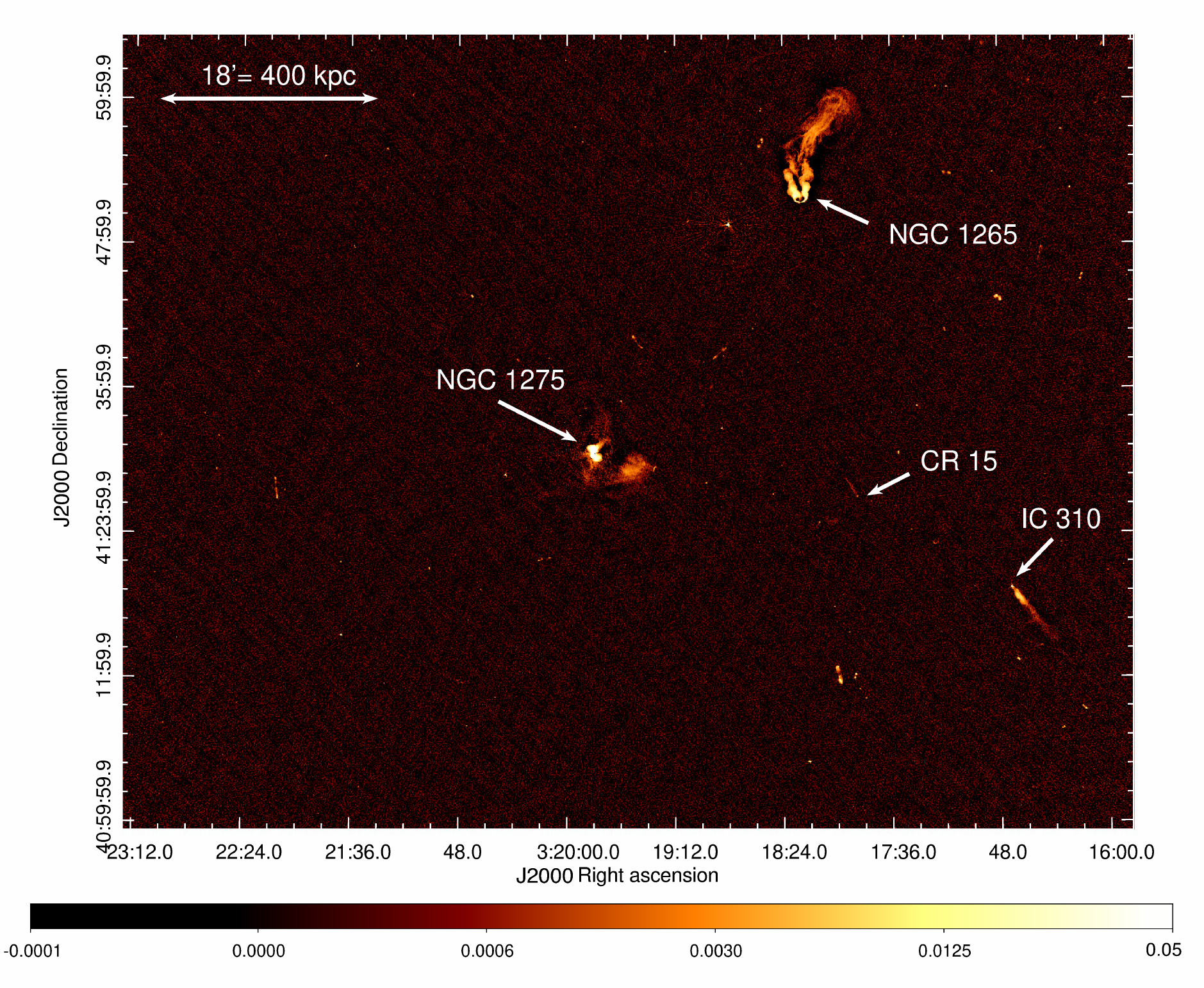}
\caption{The central $1.4\,^{\circ} \times 1.1\,^{\circ}$ of the total field of view of the VLA 230-470 MHz radio map obtained in A-configuration. Color scale units are Jy beam$^{-1}$. NGC 1275 is the bright source in the middle of the image. The wide-angle tail radio galaxy NGC 1265 (NNW of NGC 1275) and the head-tail CR 15 (between NGC 1275 and IC 310), as well as IC 310 (WSW of NGC 1275) are clearly visible. The resulting image has a rms noise of 0.27 mJy beam$^{-1}$, a beam size of $3.7 \arcsec \times 3.6 \arcsec$ and a peak of 7.34 Jy beam$^{-1}$.}
\label{fig:image_JVLA} 
\end{figure*}
\end{center}

\begin{figure*}
\begin{subfigure}{.49\textwidth}
\centering\includegraphics[height=8cm]{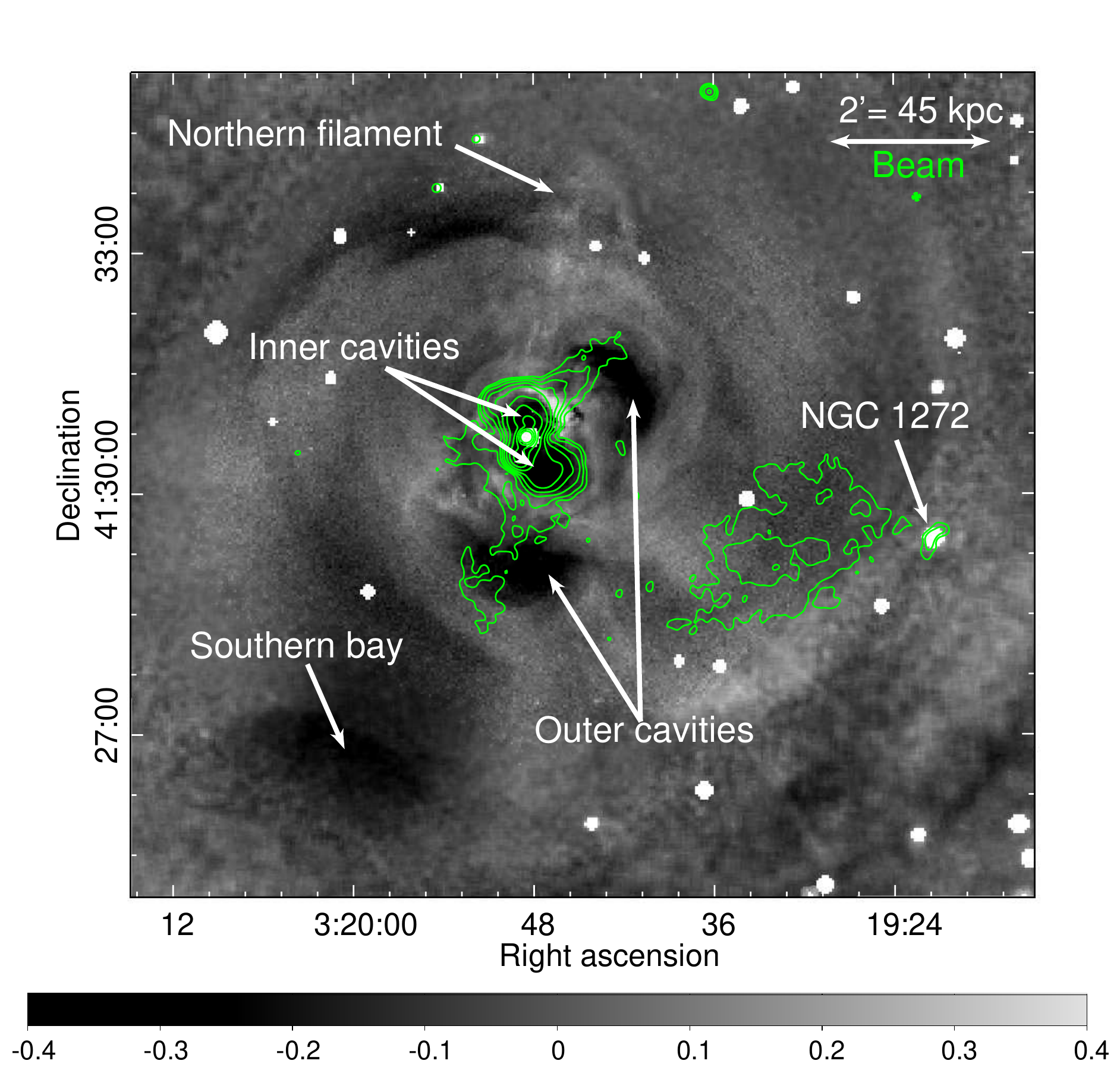}
\end{subfigure}
\begin{subfigure}{.49\textwidth}
\centering\includegraphics[height=7.6cm]{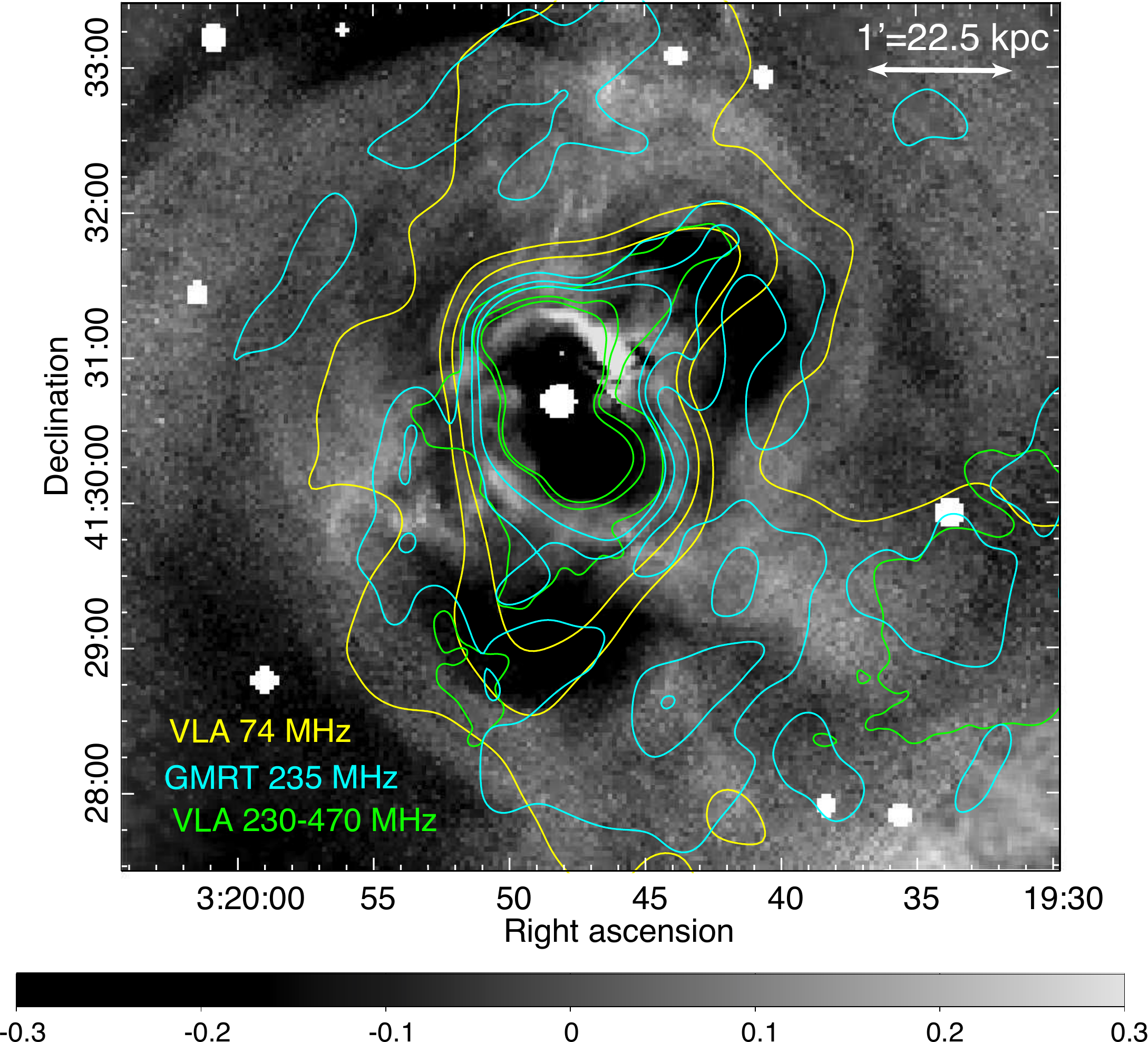}
\end{subfigure}

\vspace{-0.2cm}
\includegraphics[height=8.5cm]{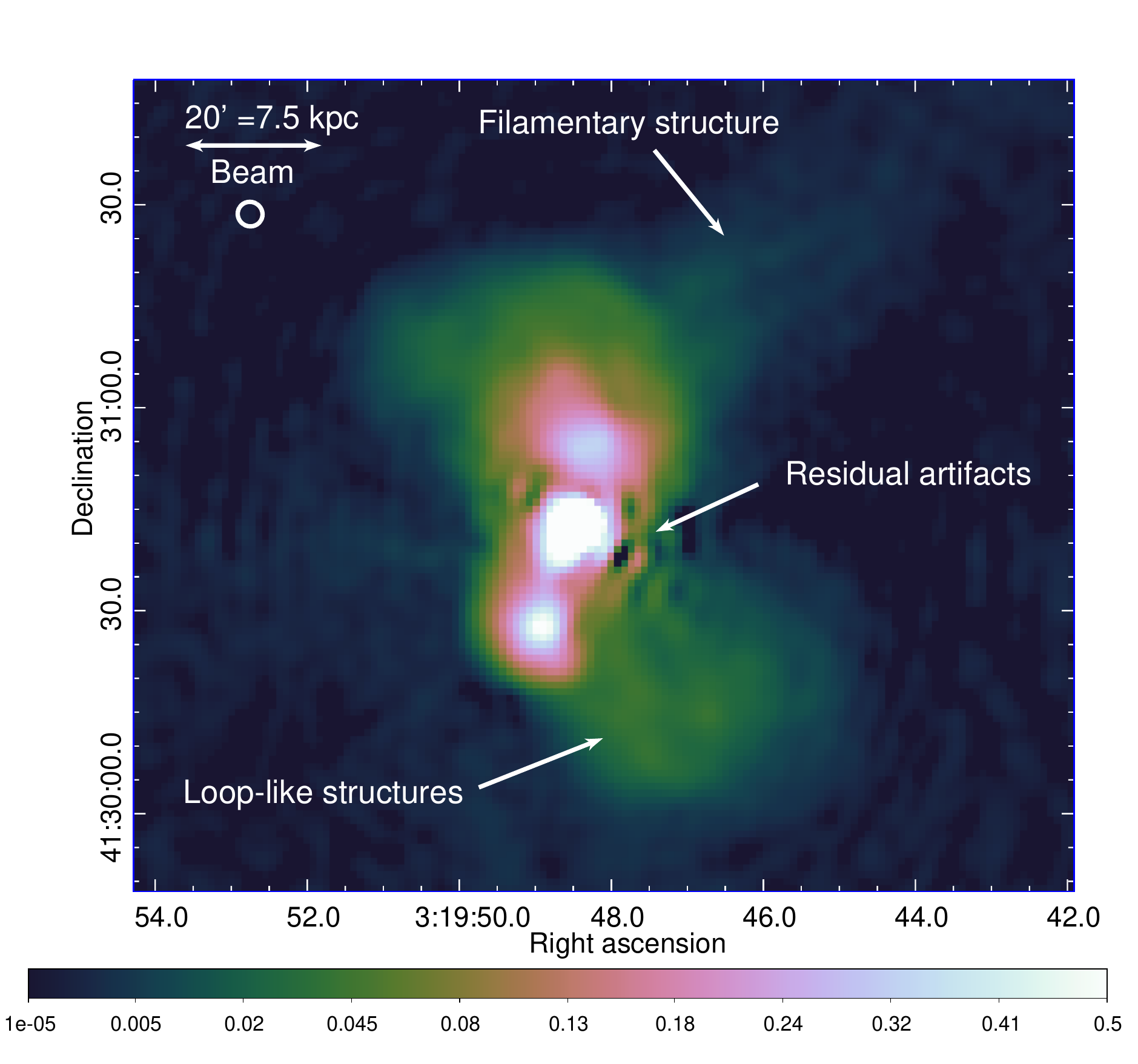}
\caption{Three images of the Perseus cluster at different scales. 
Top-left: Large-scale \textit{Chandra} final composite fractional residual image from \protect\cite{fabian_wide_2011} in the 0.5-7 keV band (1.4 Ms exposure) with A-configuration VLA 270-430 MHz logarithmic contours from $3\sigma = 0.81$ mJy beam$^{-1}$ to 1 Jy beam$^{-1}$ overlaid (10 contours levels are shown, the beam size is shown on the top-right corner). Several X-ray structures are identified: the inner and outer cavities, the northern filaments as well as the southern bay. The wide-angle tail radio galaxy NGC 1272 is also identified.
Top-right: The same \textit{Chandra} image, but zoomed-in and with radio contours at three different frequencies overlaid. In all cases, a total of 3 contours are drawn, increasing linearly from three times the noise level to 1, 0.1 and 0.01 Jy beam$^{-1}$ at 74, 235 and 270-430 MHz respectively. 74 MHz A-configuration VLA contours from \protect\cite{blundell_3c_2002} (synthesized beamwidth of $\theta_{\rm FWHM} = 24\arcsec$, $\sigma_{\rm rms} = 80$ mJy beam$^{-1}$) are shown in yellow. 235 MHz GMRT contours from \protect\cite{gendron-marsolais_deep_2017} (synthesized beamwidth of $\theta_{\rm FWHM} = 13\arcsec$, $\sigma_{\rm rms} = 10$ mJy beam$^{-1}$) are shown in light blue. Finally, our A-configuration VLA 270-430 MHz contours are shown in green.
Bottom: Zoom in on the central emission surrounding NGC 1275 from our A-configuration 270-430 MHz radio map. Color scale units are Jy beam$^{-1}$. A double filamentary structure is seen toward the northwest. The beam size is shown on the top-left corner. A few residual artifacts are visible next to the bright core, to the right. The inner lobes show patchy/filamentary rather than uniform emission, with loop-like structures visible in the southern lobe.}
\label{fig:image_chandra_JVLA_zoom_zoom} 
\end{figure*}

Similar to the B-configuration observations, we adopted a self-calibration method (amplitude and phase) for the imaging process. A first image was produced with the task \textsc{clean} \citep{hogbom_aperture_1974}. Gain corrections for amplitudes and phases were derived with \textsc{gaincal}, visually inspected and applied  with \textsc{applycal}. This procedure was applied three times. Self-calibrating the phases also allowed to correct for the TEC (Total Electron Content) variations. \textsc{bandpass} and \textsc{blcal} (baseline-based, frequency-independent, constant gain calibration, i.e. with the solution interval set to infinite) calibrations were then conducted based on the target field, to produce the final image. We used the multi-scale and multi-frequency synthesis-imaging algorithm of the task \textsc{clean} (MS-MFS, \citealt{rau_multi-scale_2011}) with 2 Taylor's coefficients to model the spectral structure of the radio emission across the broadband receivers of the VLA $P$-band.
For the wide field images produced at these frequencies, if the sky is treated as a two-dimensional plane it will introduce increasingly severe distortions around sources at increasing distance from the center.
We therefore used the W-projection algorithm \textsc{widefield} to correct these effects \citep{cornwell_noncoplanar_2008}. We used 480 w-planes, as this number was a compromise between the limited computational resources and accuracy. The size of the image was set to $12288 \text{ pixels} \times 12288 \text{ pixels}$ ($\sim 3.41\,^{\circ} \times 3.41\,^{\circ}$). The full width at half power of the field of view in the middle of $P$-band is $2.4\,^{\circ}$. We choose a cell size of $1 \arcsec \times 1 \arcsec$ based on the standard assumption for the synthesized beamwidth at 230-470 MHz with A-configuration ($\theta_{\rm FWHM} \simeq 5.6 \arcsec$ ). We used a multi-scale  \textsc{clean}-ing algorithm \citep{cornwell_multiscale_2008} to probe the different scales of the structures we were imaging: 
the point sources ($0 \arcsec$), 
the inner cavities ($15 \arcsec$), 
and the diffuse emission from the mini-halo and NGC 1265 ($30 \arcsec$, $60 \arcsec$ and $150 \arcsec$).
The tool \textsc{PyBDSF} (Python Blob Detector and Source Finder, \citealt{mohan_pybdsm:_2015}) was used to build a  \textsc{clean}-ing mask.
Wideband primary beam corrections were calculated with the CASA task \textsc{widebandpbcor}. The effective frequency at which the image was produced is 352 MHz.
The final image resulting from this data reduction process is presented in Figure \ref{fig:image_JVLA}. It has an rms of 0.27 mJy beam$^{-1}$ and beam size of $\theta_{\rm FWHM} = 3.7 \arcsec \times 3.6 \arcsec$.

\subsection{Other radio observations}

To perform a spectral index analysis of both NGC 1265 and IC 310, we used Giant Metrewave Radio Telescope (GMRT) observations at 610 MHz of these two sources presented in \cite{sebastian_giant_2017}. The calibrated measurement set was re-imaged with a $\theta_{\rm FWHM} = 15 \arcsec$ resolution and a cell size of $3 \arcsec$, in order to get similar resolution to our B-configuration VLA  270-430 MHz radio map. A minimum UV cutoff of $0.22 k\lambda$ was also applied in order to match the one of the VLA image. We corrected the position offset of $\lesssim 1 \arcsec$ between both observations. We then applied the CASA task \textsc{imregrid} to regrid the GMRT image onto the VLA one. Finally, we used the CASA task \textsc{imsmooth} to smooth both GMRT and VLA images to a common resolution, a circular beam of $\theta_{\rm FWHM} = 23 \arcsec$. When producing the spectral index map, we excluded pixels with flux densities lower than $3 \sigma_{rms}$ based on the noise level of the GMRT image (since it was higher than the VLA images).

In order to confirm features seen in the $P$-band observations, we also compare our observations with unpublished VLA data targeting IC 310, observed at $C$-band (4--8~GHz) in the C-configuration on 23 Feb 2016 as part of project 16A-291.  IC310 was observed with an on-source time of 3~min. The calibration and imaging of those data were done using CASA, version 5.1.1. Observations of 3C48 were included to set the absolute flux density scale to an accuracy of 3\% and calibrate the bandpass and delays.  For phase-referencing, we observed 3C84.   
The $C$-band data were flagged and calibrated using the VLA CASA calibration pipeline\footnote{ \href{https://science.nrao.edu/facilities/vla/data-processing/pipeline}{https://science.nrao.edu/facilities/vla/data-processing/pipeline}}. Imaging was performed with the CASA task TCLEAN using the multi-term multi-frequency-synthesis deconvolution algorithm with nterms=2 and deconvolution scales of 0, 5 and 15 pixels, 128 $w$-projection planes, Briggs weighting with robust=0.5, and a cell size of $0.6\arcsec$.  In order to mitigate deconvolution errors due to the presence of bright off-axis sources located outside the VLA $C$-band field-of-view of $\sim7.5^{\prime}$, we formed large images with dimensions of 8000 $\times$ 8000 pixels.  
In order to correct residual calibration errors, we performed a few rounds of shallow imaging and phase-only self calibration before producing our final image and applying a primary beam correction.

The final $C$-band image has an rms noise of $\sigma_{\rm rms} =$ 35~$\mu$Jy beam$^{-1}$ and synthesized beam dimensions of $\theta_{\rm FWHM} = 3.4 \arcsec\times 3.0 \arcsec$ (see Figure \ref{fig:image_perseus_ic310_zoom} - left). We corrected the position offset of $\sim 4 \arcsec$ between these observations and our A-configuration final image.
This offset could be caused by an insufficient number of W-planes for the imaging of the $P$-band data, as IC 310 is far from the pointing center (at $\simeq 37 \arcmin \simeq 820$ kpc from NGC 1275), and possibly from position shifts during self-calibration of both datasets. 


Finally, we have made use of data from the commensal VLA Low-band Ionosphere and Transient Experiment (VLITE, \citealt{clarke_commensal_2016}). NGC 1275 is used regularly at the VLA as a calibrator at high frequencies and the commensal VLITE system records the 320-384 MHz data for each of these calibration scans. Since 2017 September 01, the VLITE imaging pipeline has processed approximately 160 hours of data in more than 500 images of this field at resolutions better than $15\arcsec$. We have combined 15 pipeline images selected for image quality, rms, and resolution to create a deep image for comparison with the $P$-Band and $C$-Band images of IC 310. The VLITE images were convolved to one of two common resolutions ($5.4\arcsec$ and $14.3\arcsec$), images at each resolution were averaged, and then the two average images were feathered together. The artifacts present in the individual VLITE images are signficantly reduced by combining multiple datasets together in this way. The final image includes 30 hours of data, has a resolution of $\theta_{\rm FWHM} = 5.4 \arcsec \times 5.4  \arcsec$ and a rms noise of 0.3 mJy beam$^{-1}$ (see Figure \ref{fig:image_perseus_ic310_zoom} - middle).

\subsection{X-ray observations}

In order to compare our radio observations with the X-ray structures of the intracluster medium, we use  the final composite fractional residual image from \cite{fabian_wide_2011}, consisting of a total of 1.4 Ms \textit{Chandra} observations (900 ks of ACIS-S observations combined with 500 ks of ACIS-I observations).
The data reduction details (flare removal, reprocessing, merging, background and exposure map correcting) are shown in \cite{fabian_very_2006,fabian_wide_2011}.
Elliptical contours were fitted to the adaptively smoothed image (with a top-hat kernel of 225 counts bins) to logarithmic equally-spaced levels of surface brightness. A model was then constructed, interpolating between these contours. The fractional difference between the adaptively smoothed image and this model is shown in Figure \ref{fig:image_chandra_JVLA_zoom_zoom}. 

We also used ACIS-I \textit{Chandra} observations including the source IC 310 in their field of view: IDs 5597, 8473, and 9097, taken in FAINT mode (ObsID 5597) or VFAINT mode (ObsID 8473 \& 9097). The data were first reprocessed in Ciao 4.11 with CALDB 4.8.2 using the CIAO script \textsc{chandra\_repro} with the \textsc{check\_vf\_pha}$=$\textsc{yes} option for the latter 2 observations (i.e. those taken in VFAINT mode). Then, \textsc{wavdetect} was run on the flux corrected images created with \textsc{fluximage} to find and exclude point sources, and the \textsc{deflare} procedure was used to clean the events files of any flaring, excluding the point sources detected. Finally, these deflared events files were used with \textsc{merge\_obs} to create merged, flux-corrected images in multiple bands.  The broadband image covers the main energy range of \textit{Chandra}, 0.5-7 keV.

\section{Results}\label{Results_highres}

\begin{figure*}
\includegraphics[height=8cm]{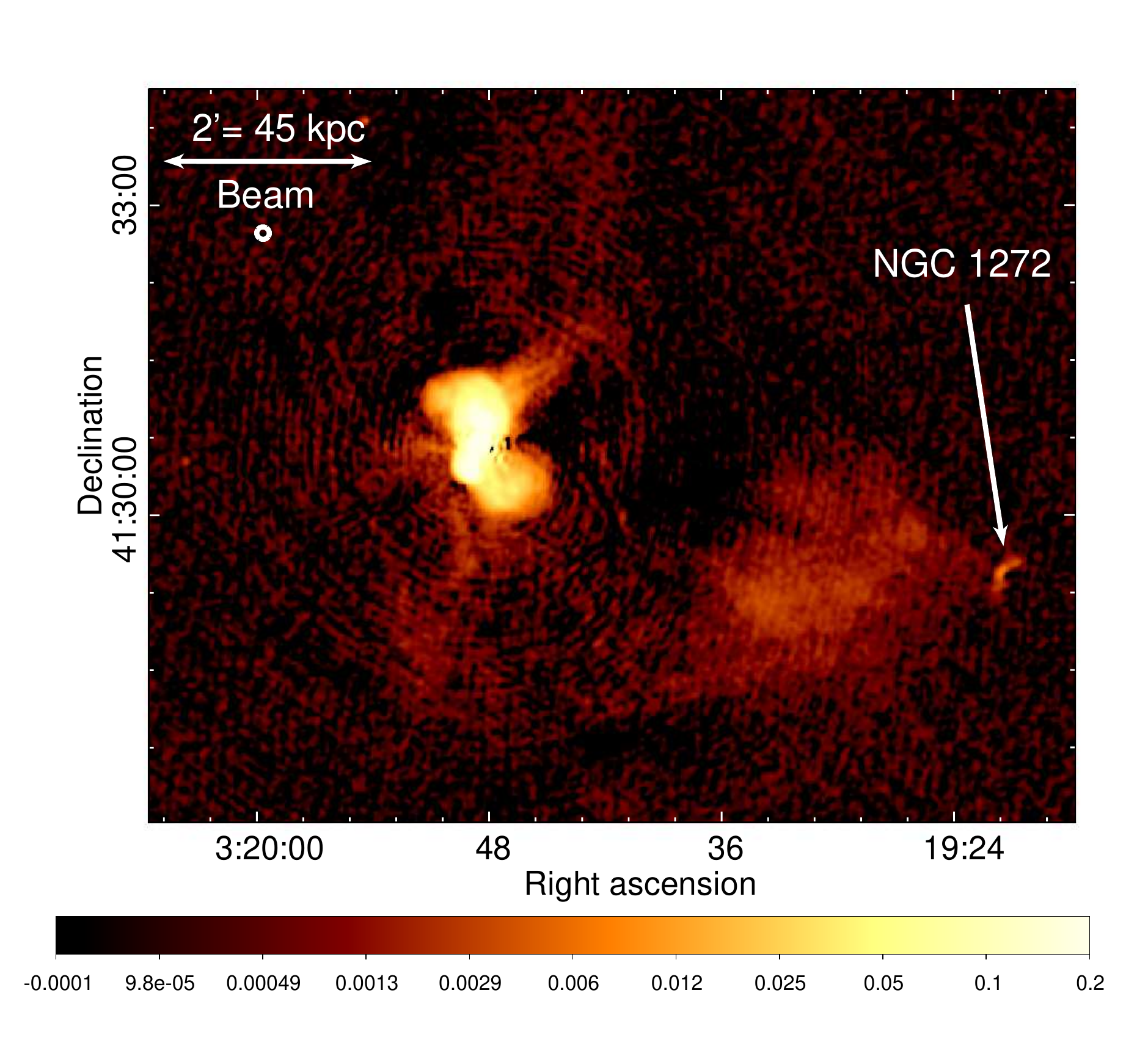}
\includegraphics[height=8cm]{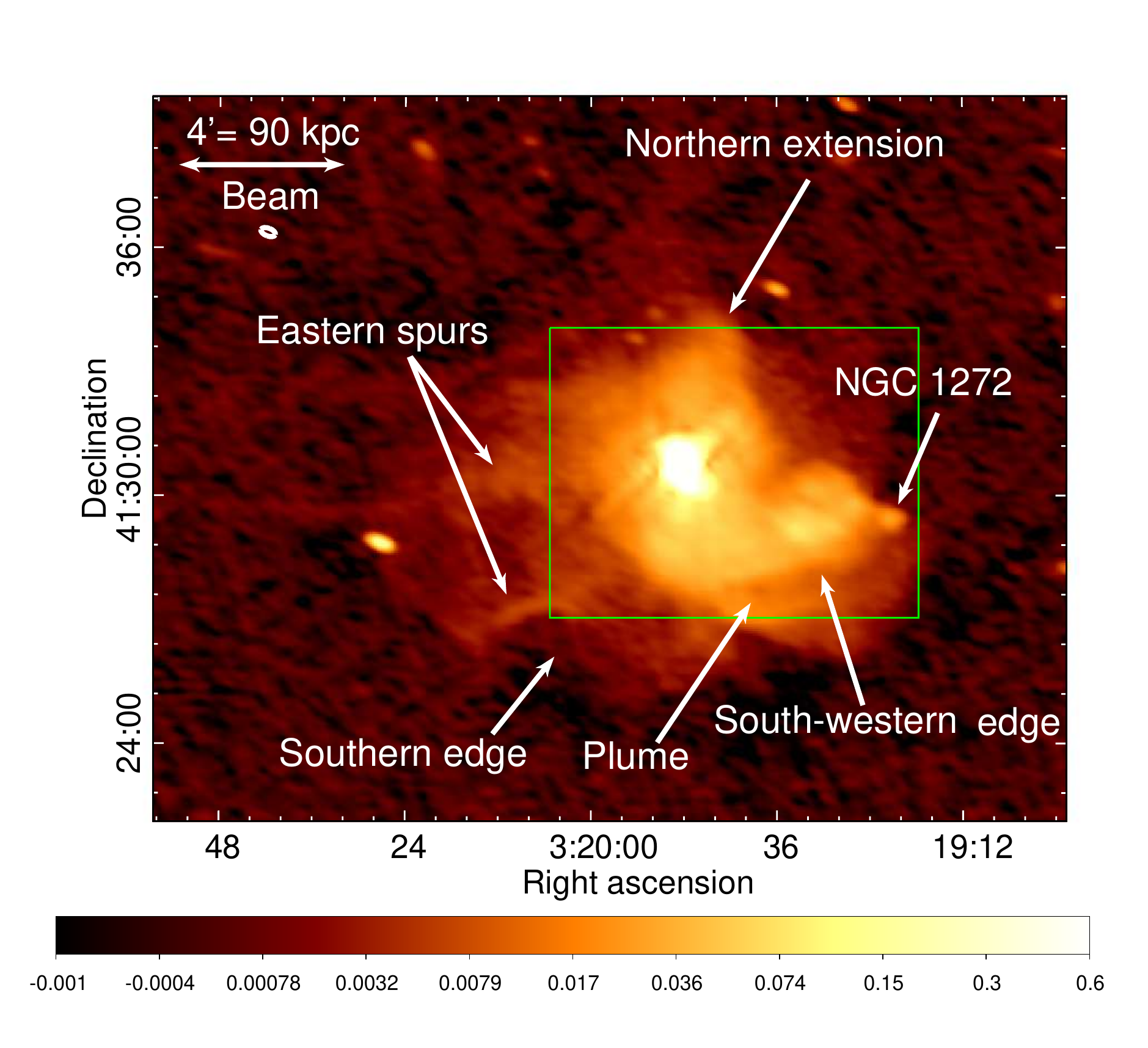}
\caption{Color scales units are Jy beam$^{-1}$. Left: zoom-in on the radio emission surrounding NGC 1275 from our A-configuration 270-430 MHz radio map shown in Figure \ref{fig:image_JVLA}. The wide-angle tail radio galaxy NGC 1272 is identified. Most of the extended emission from the mini-halo disappears at this resolution.
Right: The mini-halo emission surrounding NGC 1275 from the B-configuration 270-430 MHz radio map from \protect\cite{gendron-marsolais_deep_2017}. The green square shows the size of the A-configuration image on the left. The main structures of the mini-halo are identified: the northern extension, the two eastern spurs, the concave edge to the south, the south-western edge and a plume of emission to the south-south-west. The small knob at the end of the western tail is NGC 1272. This image has a rms noise of 0.35 mJy beam$^{-1}$, a beam size of $\theta_{\rm FWHM} = 22.1 \arcsec \times 11.3 \arcsec$ and a peak of 10.63 Jy beam$^{-1}$.}
\label{fig:image_JVLA_zoom} 
\end{figure*}

Figure \ref{fig:image_JVLA} shows the central part of the A-configuration final map obtained from the data reduction and image processing described in Section \ref{VLA Observations and data reduction_highres}. 
We reach a  dynamic range of $27 000:1$, with an rms of 0.27 mJy beam$^{-1}$ and a peak at 7.34 Jy beam$^{-1}$ that coincides with the AGN in NGC 1275. The beam size is $\theta_{\rm FWHM} = 3.7 \arcsec \times 3.6 \arcsec$ and is shown on the upper-left corner of the Figure \ref{fig:image_JVLA_zoom} - right.
This is similar to the dynamic range reached for the VLA B-configuration observations at the same frequencies, that is $30 000:1$ (rms of 0.35 mJy beam$^{-1}$ and a peak at 10.63 Jy beam$^{-1}$).
It is also almost four times deeper than the previous A-configuration VLA observations of Perseus, which reached a dynamic range of $7300:1$ (rms of 1.2 mJy beam$^{-1}$ and a peak at 8.79 Jy beam$^{-1}$, \citealt{blundell_3c_2002}).
Observations from the WRST at 327 MHz reached a dynamic range of $16 500:1$ (rms of 1.4 mJy beam$^{-1}$, \citealt{sijbring_radio_1993}) and have a much larger beam ($\theta_{\rm FWHM} = 51 \arcsec \times 77 \arcsec$).

The radio emission filling the core of the Perseus cluster can be divided in two components: the radio lobes corresponding to the cavity system seen in X-ray observations and the diffuse mini-halo emission. Both will be described in Section \ref{NGC 1275_highres}.
Beyond the central emission from NGC 1275, the large field of view also includes several complex radio sources, including IC 310, NGC 1265, CR 15 and NGC 1272. Their morphologies and spectral index distributions will be described and analyzed in Sections \ref{Head-tail sources in the Perseus cluster_highres} and \ref{Spectral index maps of NGC 1265 and IC 310}.

\section{NGC 1275}\label{NGC 1275_highres}

\subsection{High-resolution observations of the radio lobes} \label{radio lobes_highres}

Radio emission at 230-470 MHz from our A-configuration observations completely fills the inner X-ray cavities as shown in Figure \ref{fig:image_chandra_JVLA_zoom_zoom}. New hints of sub-structures appear in the inner lobes, so that the inner cavities are not filled with smooth uniform radio lobes. Instead, they contain patchy/filamentary emission, similar to the web of thin filaments seen inside the radio lobes of M87 \citep{owen_m87_2000}. New loop-like structures are visible in the southern lobe. Higher resolution observations are however required to resolve these sub-structures. A few residual artifacts are visible next to the core, to the west. 
The southern inner lobe is clearly made of two parts: the bright one, closer to the nucleus, and oriented towards the south-east and a fainter part, oriented towards the south-west. This two-part structure has been  known since the first VLA observations of NGC 1275 \citep{pedlar_radio_1990}. 

Furthermore, fainter spurs of radio emission extend into both of the outer cavities. These were first discovered at 74 MHz in \cite{blundell_3c_2002} with the VLA and later detected at 235 MHz with the GMRT \citep{gendron-marsolais_deep_2017}. 
With our new $P$-band image, it is now the first time they are seen at higher frequency ($>235$ MHz). Our higher-resolution VLA observations also allow us to resolve the northwestern spur into a double filamentary structure (see Figure \ref{fig:image_chandra_JVLA_zoom_zoom}, bottom). These spurs have not been detected at frequencies higher than $\sim 500$ MHz due to a lack of published observations with appropriate resolution and sufficient depth. If improved observations above 500 MHz still do not detect the emission in the outer cavities, this would be consistent with a spectral turnover at around 350 MHz, which is the spectral behavior expected from the traditional spectral aging interpretation of the emission in radio lobes. Indeed, for a magnetic field of about $10 \mu \text{G}$, a spectral turnover at around 350 MHz implies that the synchrotron age of the particle distribution is $\sim 70$ Myr, in agreement with the buoyancy rise time of the northern ghost cavity ($\sim 60$ Myr, see \citealt{fabian_properties_2002}). Overlaying the 74 MHz, 235 MHz and 270-430 MHz contours onto the \textit{Chandra} observations (see Figure \ref{fig:image_chandra_JVLA_zoom_zoom}, top-right), we show how the emission into the outer lobes is more extended as the frequency is lower, as expected from spectral aging. However, this might also be due to the resolution of the observations, which decreases with frequency, catching more and more of the diffuse mini-halo emission and making the comparison between the images hard to quantify. Indeed, the largest angular scales recoverable in the 74 MHz and 235 MHz observations ($\sim 15\arcmin$ and $\sim 45\arcmin$, respectively) are larger than in the $P$-band A-configuration observations ($\sim 3\arcmin$). Unfortunately, there are currently no observations below $P$-band frequencies with sufficiently high spatial resolution to disentangle the jet emission from the mini-halo.

\subsection{High-resolution observations of the mini-halo} \label{mini-halo_highres}

The Perseus cluster is known to host a mini-halo \citep{soboleva_3c84_1983,pedlar_radio_1990,burns_where_1992,sijbring_radio_1993}, its diffuse emission filling the central $\sim200$ kpc of the cluster core. 
VLA $P$-band B-configuration observations of Perseus have revealed a rich variety of complex structures in this mini-halo (see Figure \ref{fig:image_JVLA_zoom}, \citealt{gendron-marsolais_deep_2017}). Almost all of this extended emission disappears in the higher resolution VLA observations as the largest angular scale recoverable with the A-configuration is $\sim 3 \arcmin$ at  $P$-band frequencies. 
Only some of the brightest western parts of the mini-halo remain. From our image, it is not clear if this diffuse emission is related to the wide-angle tail radio galaxy NGC 1272 (see Section \ref{NGC 1272_highres}). Deeper, higher-resolution imaging at $\lesssim 1$ GHz with spectral studies of this emission are needed to attest this potential link. Assuming that mini-halos originate from the reacceleration of pre-existing electrons by turbulence \citep{gitti_modeling_2002,gitti_particle_2004}, if part of the mini-halo is indeed linked to the tail of NGC 1272, it would suggest that the mini-halo emission is not all coming from the reacceleration of the AGN particles, originating from NGC 1275, but rather that some of this emission is generated by the fossil population of particles released by NGC 1272. This would be very similar to what is observed in the merging cluster Abell 3411-3412, where a clear connection is seen between a narrow-angle tail radio galaxy and a radio relic \citep{van_weeren_case_2017}. Radio galaxies may therefore also play a role in the diffuse non-thermal emission found in relaxed clusters such as mini-halos emission. For now, there is no clear evidence of this connection in relaxed cool-core clusters, although similar radio structures are seen in the cool-core cluster RXJ1720.1+2638
\citep{giacintucci_mapping_2014,savini_lofar_2019}. Indeed, at low resolution, radio observations of RXJ1720.1+2638 show a large mini-halo with a spiral-shaped structure, while at higher resolution, a head-tail radio galaxy is also visible, its tail connecting with the central diffuse emission through a channel of radio emission with steep spectra. Perseus and RXJ1720.1+2638 could be the first clear examples of connection between a bent-jet radio galaxy and mini-halo emission.

\begin{figure*}
    \centering
    \includegraphics[width=0.4\textwidth]{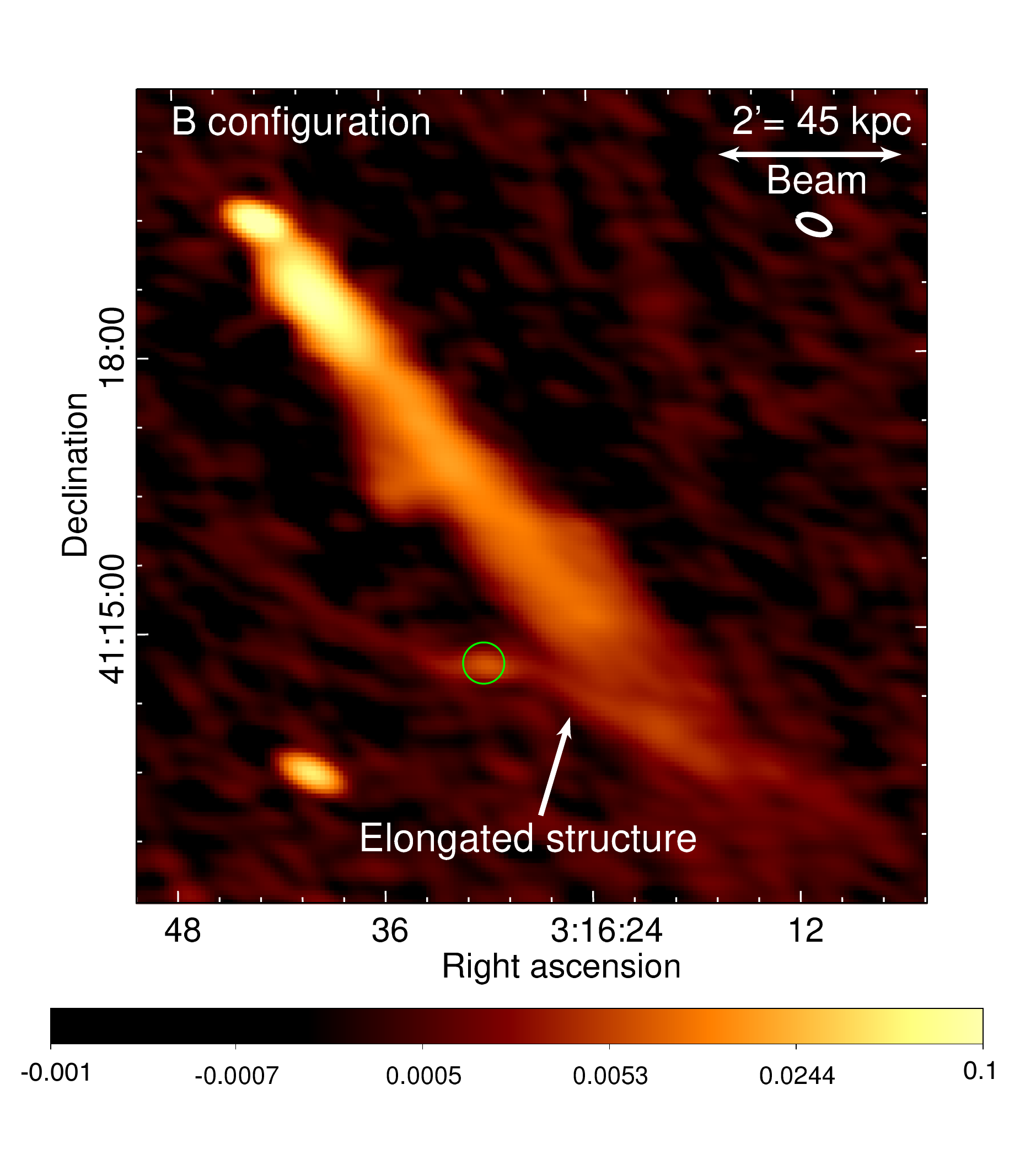}
    \includegraphics[width=0.4\textwidth]{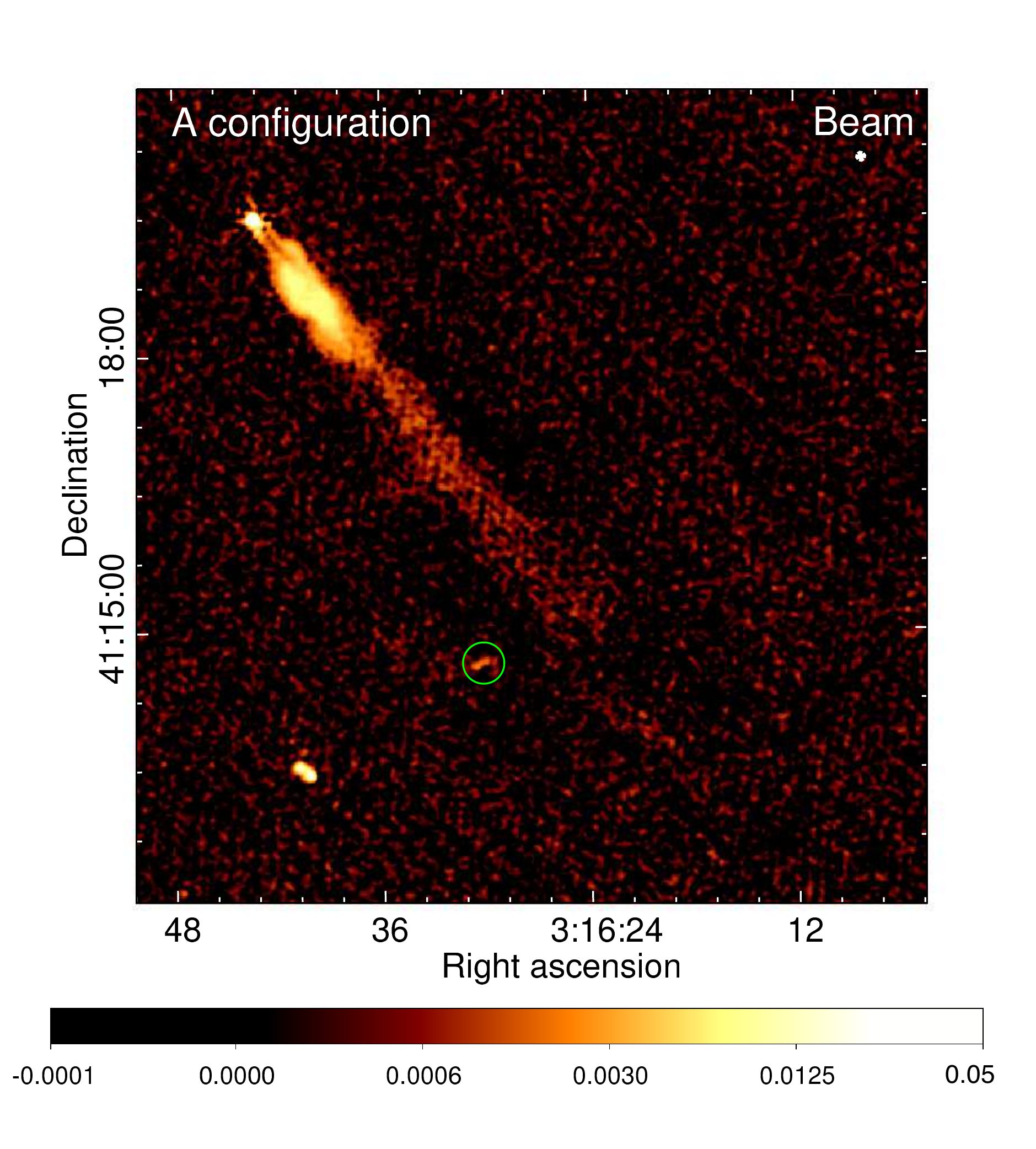}
\caption{The radio galaxy IC 310 at 230-470 MHz with two different configurations from our VLA observations. Both images are the same size and the color scales units are Jy beam$^{-1}$. Left - B-configuration observations (rms noise of 0.35 mJy beam$^{-1}$, beam size of $\theta_{\rm FWHM} = 22.1 \arcsec \times 11.3 \arcsec$). Distortions in the tail are visible as well as a strange elongated source merging with the tail. Right - A-configuration observations (rms noise of 0.27 mJy beam$^{-1}$, beam size of $3.7 \arcsec \times 3.6 \arcsec$). The green circle shows the position of the source found in the A-configuration image potentially linked to the elongated structure found in the B-configuration image.}
\label{fig:image_perseus_ic310}
\end{figure*}

\begin{figure*}
    \centering
    \includegraphics[width=0.32\textwidth]{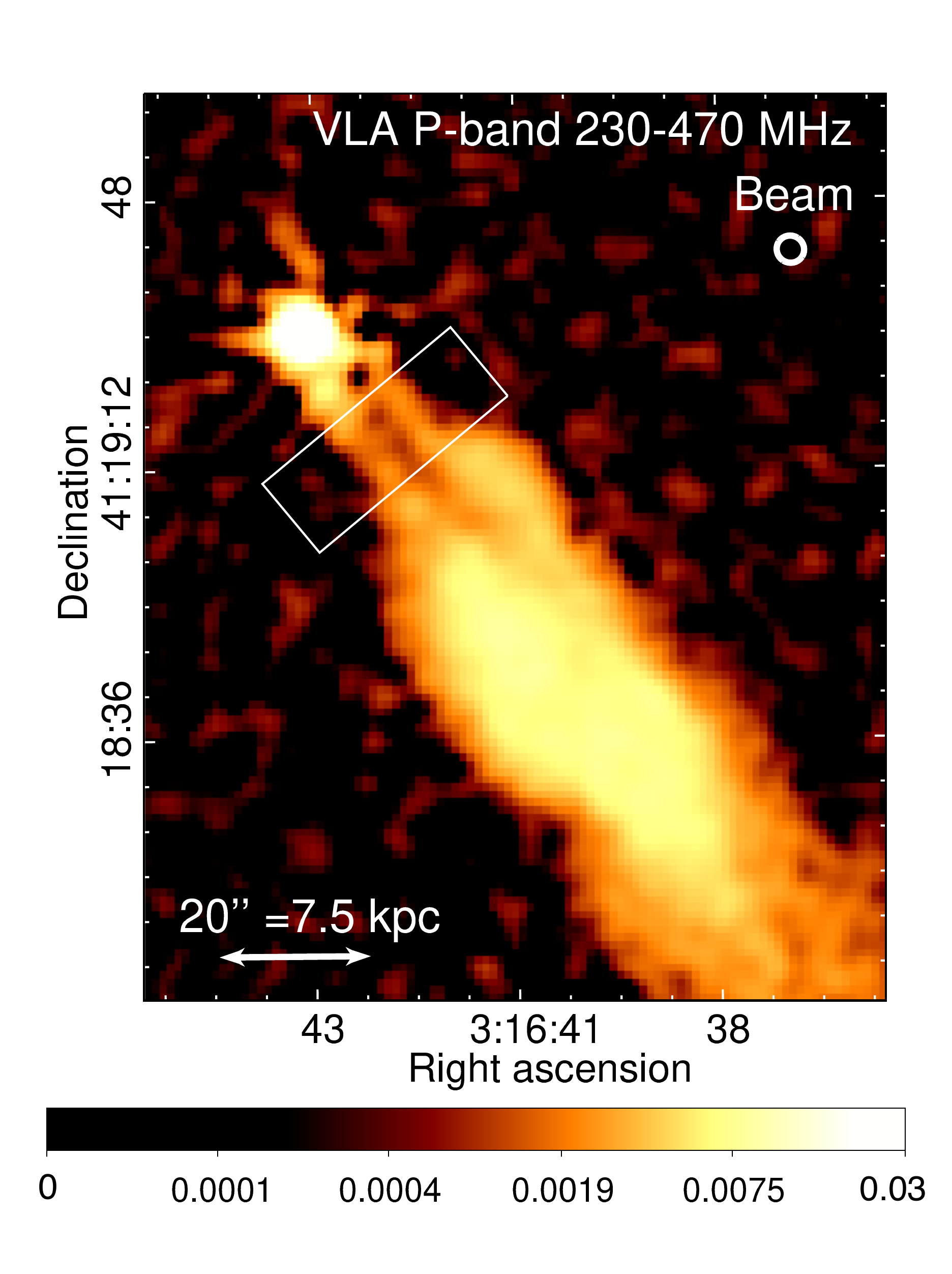}
    \includegraphics[width=0.32\textwidth]{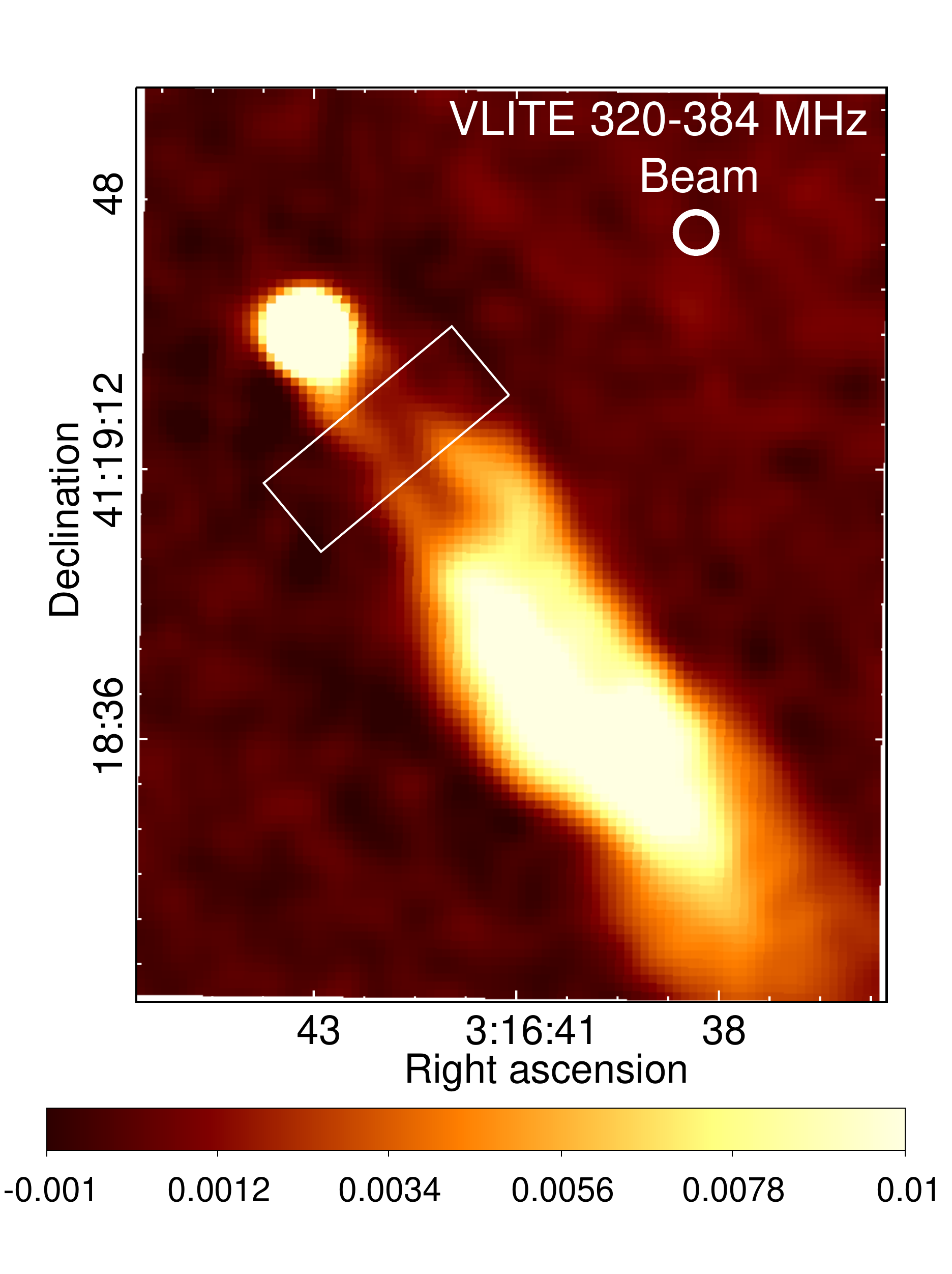}
    \includegraphics[width=0.32\textwidth]{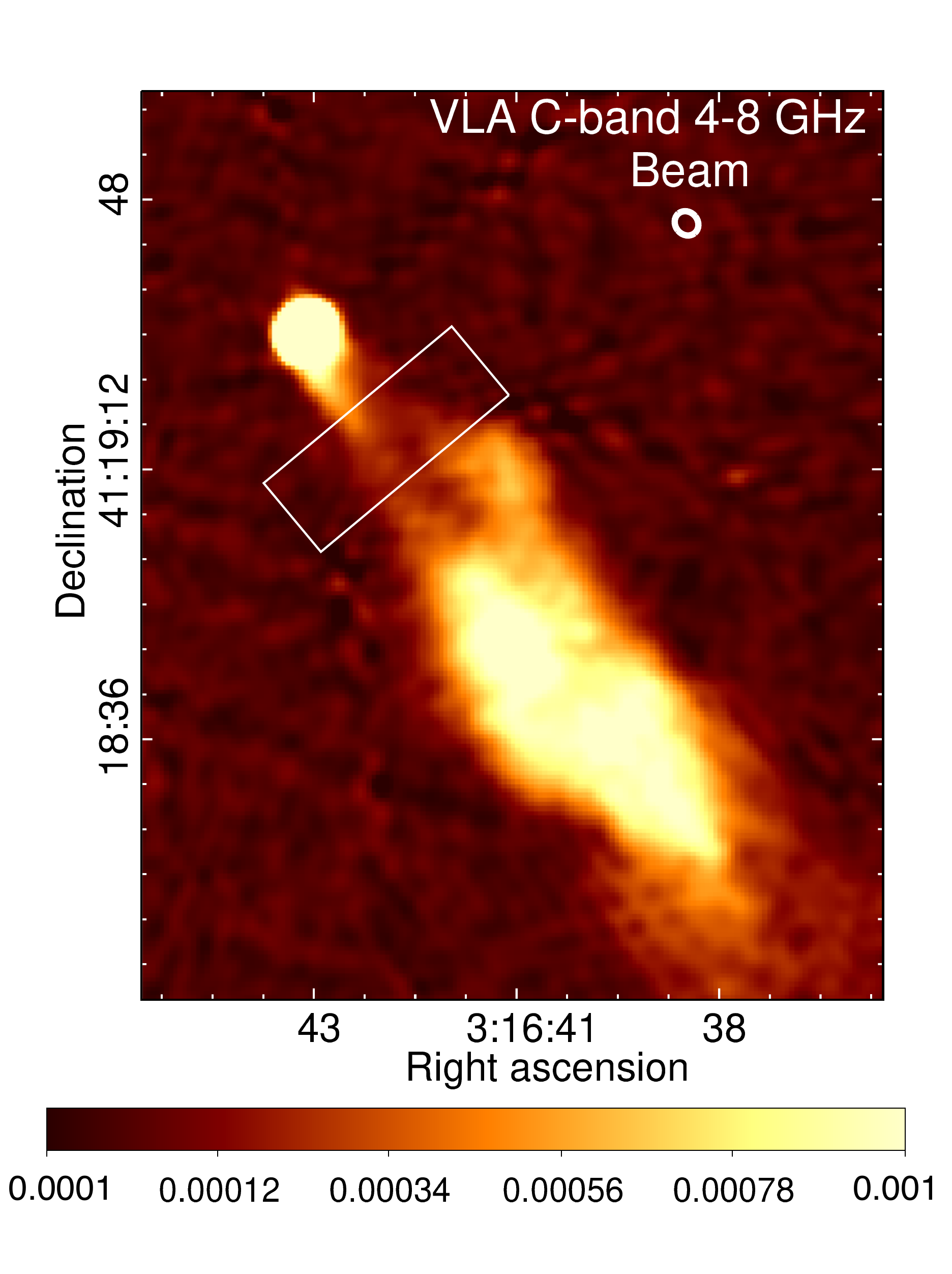}
\caption{Zoom-in on the base of the jets of the radio galaxy IC 310. All images are the same size and color scale units are Jy beam$^{-1}$. The profiles extracted from the white box across the base of the jets are shown in Figure \ref{fig:ic310_profile}. Right - Our VLA 230-470 MHz A-configuration observations show two distinguishable collimated narrow jets. Middle - VLITE 320-384 MHz image of IC 310 (rms noise of 0.3 mJy beam$^{-1}$ and beam size of $\theta_{\rm FWHM} = 5.4 \arcsec \times 5.4  \arcsec$). Despite the lower resolution, the bridge of emission between the head and the tail splits, similarly to what is seen in the $P$ and $C$ band images.
Left - The  $C$-band (4--8~GHz) VLA observations (rms noise of 0.035 mJy beam$^{-1}$ and beam size of $\theta_{\rm FWHM} = 3.4 \arcsec \times 3.0 \arcsec$) show very similar features as the low-frequency image, however the jets are less clearly distinguishable near the nucleus. }
\label{fig:image_perseus_ic310_zoom}
\end{figure*}

\begin{figure}
    \centering
    \includegraphics[width=\columnwidth]{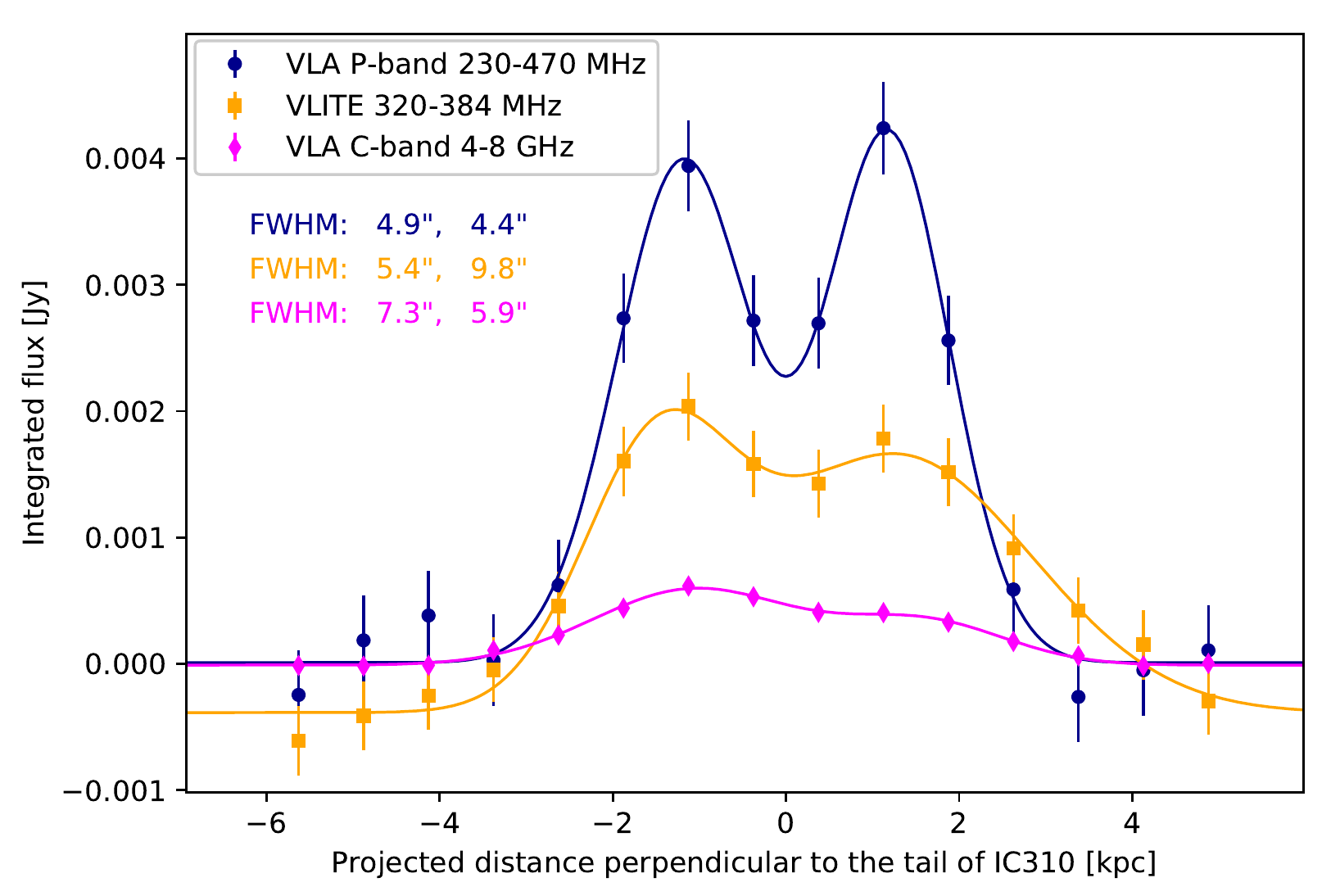}
\caption{Profiles extracted from the three images of IC 310 of Figure \ref{fig:image_perseus_ic310_zoom}  (VLA P-band A-configuration, VLITE and $C$-band) in a rectangular region ($33\arcsec\times12\arcsec$) across the base of the jets. The lines show the results of the fit with the sum of two Gaussian functions. The fitted FWHMs are shown on the plot. 
In addition to the statistical errors plotted, there is a systematic error due to possible flux scale errors of $\sim 10\%$.}
\label{fig:ic310_profile}
\end{figure}

\begin{figure*}
\centering
\includegraphics[width=1.6\columnwidth]{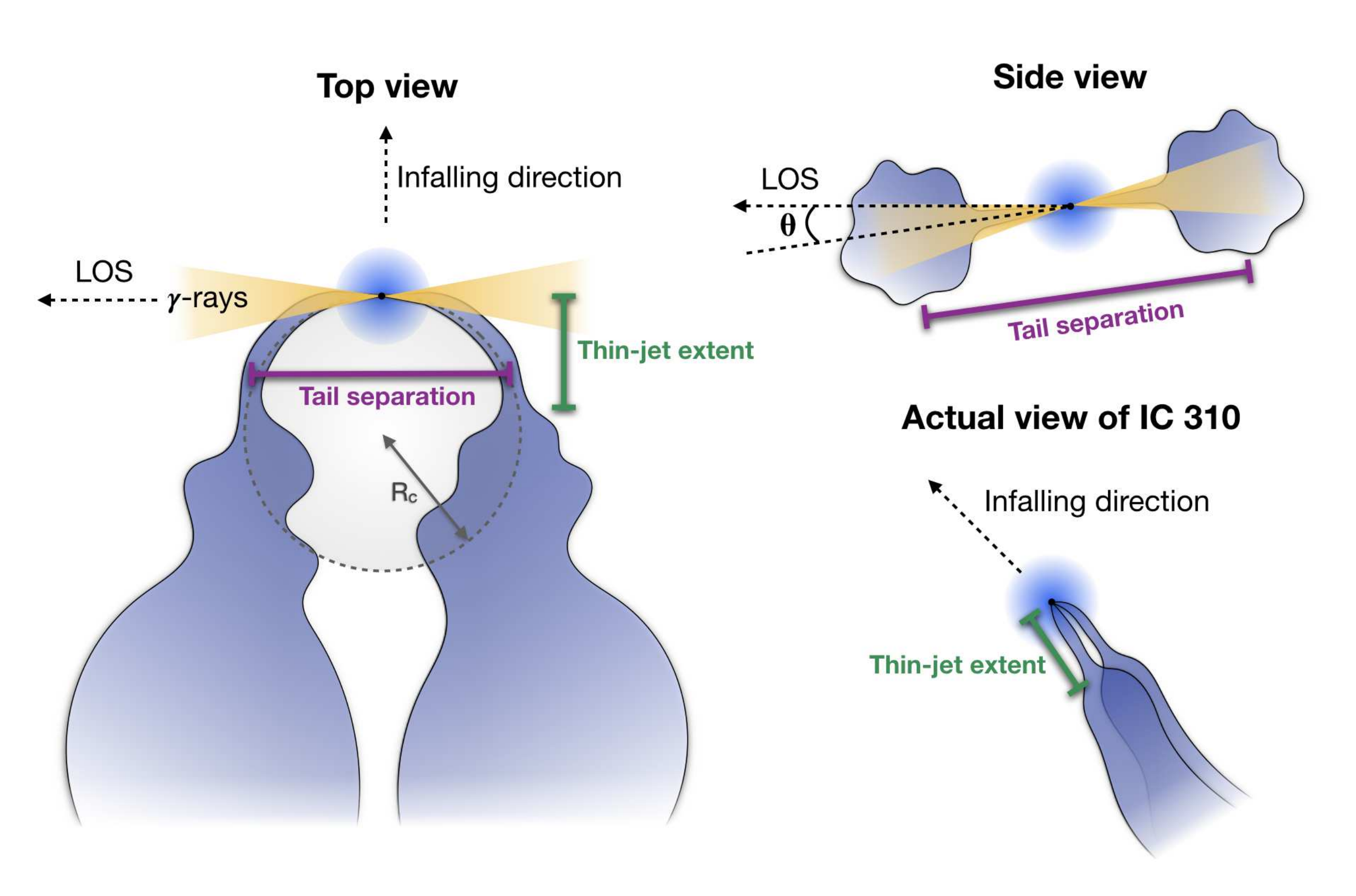}
\caption{Schematic representation of the head-tail source IC 310 from three different point of views, where the detected bifurcation is interpreted as two bent jets. Here, the jets are closely aligned with our line of sight (with an angle $\theta$) and the $\gamma$-rays are beamed toward us, instead of following the bent jets, consistent with the $\gamma$-ray detection from IC 310. According to this representation, the source can be characterized by different lengths: the tail separation, the thin-jet extent and the radius of curvature of the bent jets ($R_c$). From our line of sight, the thin-jet extent can be directly measured in IC 310.}
\label{fig:IC310_schema}
\end{figure*}

\begin{figure*}
    \centering
    \includegraphics[width=0.45\textwidth]{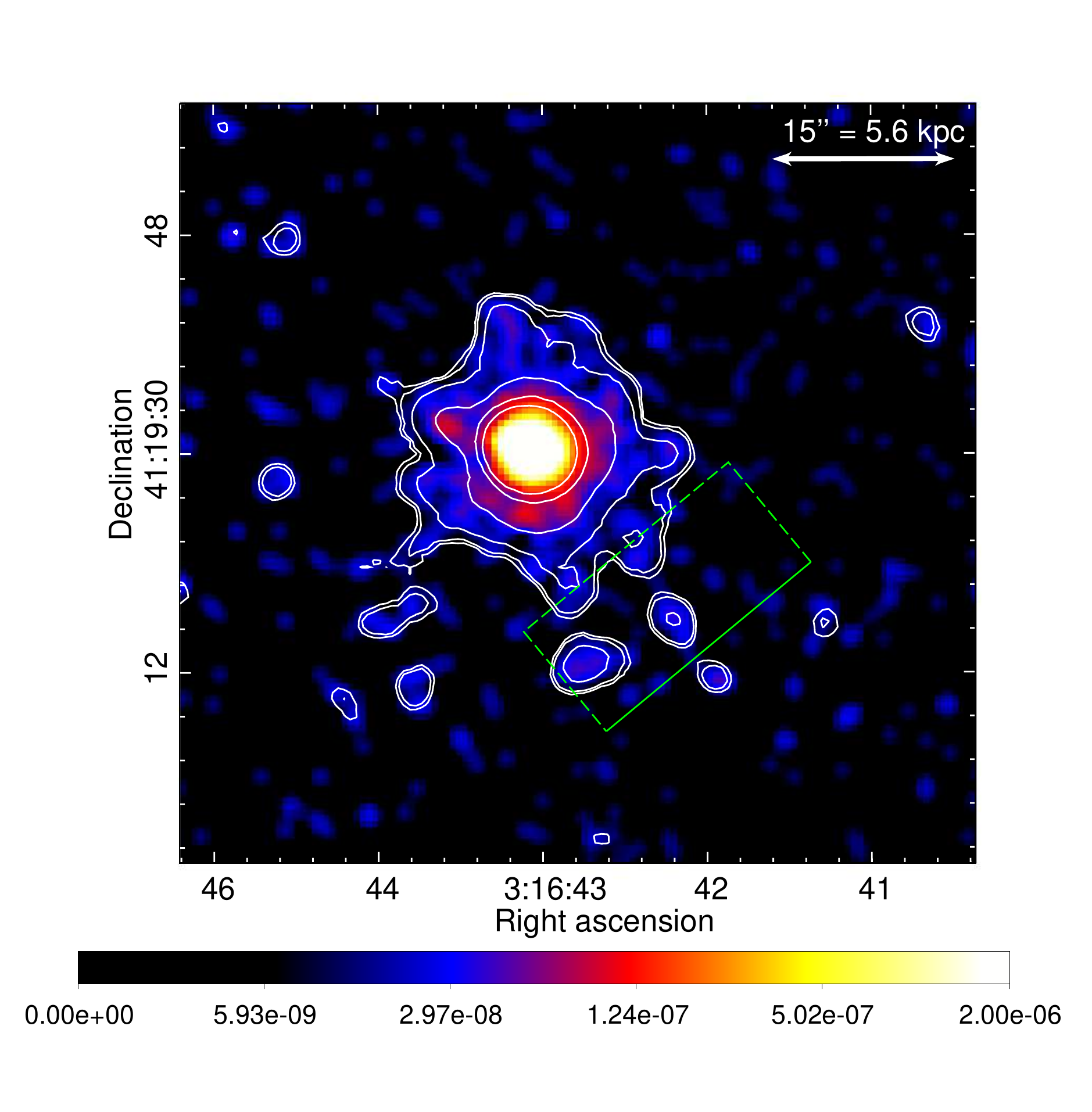}
    \includegraphics[width=0.45\textwidth]{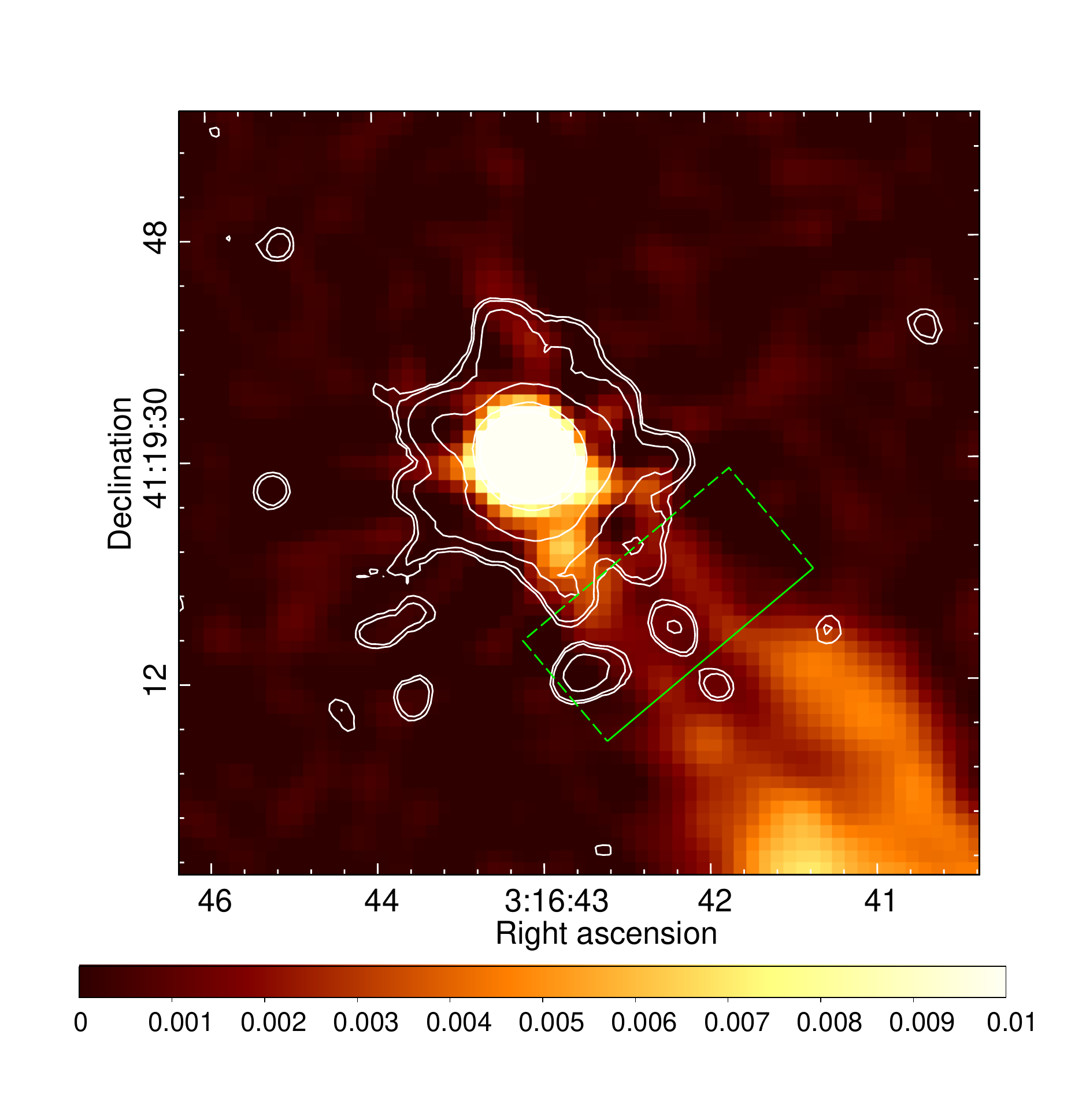}
\caption{Left - The X-ray minicorona surrounding the head-tail radio galaxy IC 310 from the \textit{Chandra} observations with $3\sigma$ contours. Color scale units are $\text{counts}/\text{s}/ \text{cm}^2$.
Right - The same field-of-view, showing our A-configuration $P$-band VLA observations with the X-ray contours overlaid. Color scale units are Jy beam$^{-1}$.}
\label{fig:image_perseus_ic310_x-ray}
\end{figure*}

\section{Bent-jet radio galaxies in the Perseus cluster}\label{Head-tail sources in the Perseus cluster_highres}

\subsection{IC 310}\label{IC310_highres}

The S0 galaxy IC 310, member of the Perseus cluster ($z=0.0189$, \citealt{miller_radio_2001,tully_cosmicflows-3_2016}) and first discovered by \cite{ryle_radio_1968}, was originally classified as a narrow-angle tail radio galaxy \citep{sijbring_multifrequency_1998} based on its morphology, showing only a one-sided tail interpreted as the fusion of two jets strongly bent by ram pressure stripping of the ICM. 
However, it was recently shown that its nucleus has a blazar-like behaviour (for a review, see e.g. \citealt{glawion_black_2016}). 
It was first detected above 30 GeV with a $6\,\sigma_{ \rm rms}$ statistical significance with the \textit{Fermi}-Large Area Telescope (LAT, \citealt{neronov_very_2010}).
Then, it was also detected above 300 GeV with the MAGIC telescopes at a  statistical significance of $7.6\, \sigma_{\rm rms}$ \citep{aleksic_detection_2010}. The very high energy $\gamma$-rays flux also shows a day-scale variability \citep{aleksic_rapid_2014}.
In 2012, the MAGIC telescope detected an exceptionally bright flare of IC 310, reaching an averaged flux level above 1 TeV of up to one Crab \citep{aleksic_black_2014}.
Causality constraints arising from the rapid variability observed during this flare have restrained the size of the $\gamma$-ray emission region to $<20\%$ of the central black hole gravitational radius \citep{aleksic_black_2014}.
More recently, variability studies with \textit{Fermi}-LAT have also reported the detection of a soft emission state, with detection of emission below 1 GeV \citep{graham_fermi-lat_2019}.
Very long baseline interferometry (VLBI) observations have also shown only one counter jet visible at parsec scales oriented in the same direction as the kpc tail \citep{kadler_blazar-like_2012,schulz_evn_2015}. 
Overall, these arguments disagree with the head-tail classification of IC 310. Instead, it could be a low-luminosity FRI radio galaxy \citep{rector_search_1999} with its jets nearly aligned along our line of sight. The angle between the plane of the two jets and our line of sight is estimated to be $10\,^{\circ} - 20\,^{\circ}$, at a borderline angle to reveal its BL Lac-type central engine \citep{kadler_blazar-like_2012,aleksic_black_2014,glawion_black_2016}. The upper limit of this estimation comes from the non-detection of the pc-scale counter-jet in VLBI observations, taking into account Doppler boosting. The lower limit on this angle is estimated assuming a maximum de-projected extent of the collimated jet of IC 310 to be 1 Mpc, considering its known projected extent of $\sim 400$ kpc 
based on the emission detected in WSRT observations down to $3\sigma_{\rm rms}$ level \citep{sijbring_multifrequency_1998}.

Nevertheless, our $P$-band observations show new radio structures in IC 310. At low-resolution (see B-configuration observations on Figure \ref{fig:image_perseus_ic310} - left), distortions in the tail are seen as well as a strange faint elongated source that merges with the end of IC 310 tail. The total projected extent of the tail measured from the $3\sigma_{\rm rms}$ level of these observations is $\sim 9 \arcmin \simeq 200$ kpc at the redshift of the cluster, while a diffuse extension of the tail of up to $\sim 400$ kpc is seen in WSRT observations of IC 310 at similar frequencies and lower resolution \citep{sijbring_multifrequency_1998}. The largest angular scale recoverable with the WSRT observations is also much larger than in the VLA $P$-band B-configuration observations ($\sim 55 \arcmin$ vs $\sim 10 \arcmin$, respectively). Surprisingly, at high-resolution (see Figure \ref{fig:image_perseus_ic310} - right and Figure \ref{fig:image_perseus_ic310_zoom} - left), two narrow collimated distinct jets are visible, attached to the nucleus. Furthermore, the long structure merging with the tail found in the B-configuration image seems to resolve into a slightly resolved source in the A-configuration image. This source could therefore be a simple radio galaxy outside of the Perseus cluster, although there does not seem to be an optical counterpart in the \textit{Digitized Sky Survey} (DSS, \citealt{lasker_digitized_1994}) or in the \textit{Panoramic Survey Telescope \& Rapid Response System} (Pan-STARRS, \citealt{kaiser_pan-starrs_2010}). Overall, Figure \ref{fig:image_perseus_ic310} - right seems to suggest that IC 310 is indeed made of two distinct jets, consistent with the original interpretation of a narrow-angle tail radio galaxy infalling into the cluster.

To confirm the presence of these structures, we examine other radio observations of IC310. However, there are very few other observations with high enough spatial resolutions and reaching low level of noise capable of resolving the region between the core and the rest of the tail.
For example, using archival 4.9 GHz VLA C-configuration observations ($\theta_{\rm FWHM} = 3.99 \arcsec \times 3.93 \arcsec$ and rms of 73.9 $\mu\text{Jy / beam} $), \cite{dunn_radio_2010} found that the tail in IC 310 is disconnected from the bright core. The authors suggest an interruption in the AGN activity, the tail being the remnant of past activity while the core presenting a more recent outburst.
We also observe a decrease in flux between the core and the tail, but both seem to remain connected by the two resolved jets. At higher frequencies, we present here the $C$-band (4--8~GHz) VLA snapshot observations of IC310 in Figure \ref{fig:image_perseus_ic310_zoom} - right. These observations show features that are remarkably similar to the low-frequency image, however the jets are less clearly distinguishable near the nucleus. Indeed, the northern jet is not as straight as what we see in the $P$-band image and its connection to the nucleus is less clear. At $C$-band, it is possible that the emission could be dominated by a different component due to the physics of particle acceleration. 
Furthermore, the presence of imaging artifacts in both images can drive our visual interpretation, as they can enhance or suppress the flux of the emission closest to the core of IC 310. However, since no other source in the field-of-view of our $P$-band A-configuration image shows a similar effect to the bifurcation seen in IC 310, we do not think this feature is an imaging artifact. Some imaging artifacts are visible in this image, for example the small linear features projecting radially from the bright core of IC 310, but these are at least a factor two smaller in size than the bifurcation. In the case of the $C$-band image, there are also some radial artifacts surrounding the core of IC 310. The tail overlaps with two artifact strikes and could therefore be affected. To further investigate the potential impact of artifacts on the apparent bifurcation, we also show the combined 320-384 MHz VLITE image we have produced in Figure \ref{fig:image_perseus_ic310_zoom} - middle, where the morphology of the bifurcation region near the core of IC 310 can be compared directly to $P$ and $C$-band images. We note that the apparent jet bifurcation matches well to that observed in the $P$-band data. We therefore conclude that this bifurcation is real.

We have extracted profiles across the base of the potential jets in IC 310 (see Figure \ref{fig:ic310_profile}) from our three images of IC 310 shown in Figure \ref{fig:image_perseus_ic310_zoom}  (VLA P-band A-configuration, VLITE and $C$-band). These profiles were fitted with the sum of two Gaussian functions. For our P-band A-configuration image, the resulting FWHMs of the two Gaussian functions fitting the jets are larger ($4.9 \pm 0.4\arcsec$ and $4.4 \pm 0.3\arcsec$) than the beam size ($\theta_{\rm FWHM} = 3.7 \arcsec \times 3.6 \arcsec$), meaning that the bifurcation is resolved. The distance between the centers of the two Gaussian functions is 2.4 kpc. We obtain similar results with the VLITE and $C$-band data, where the fitted FWHMs of the Gaussians are larger ($5.4 \pm 0.6 \arcsec$ and $9.8 \pm 1.3 \arcsec$ for VLITE, $7.3 \pm 0.5\arcsec$ and $5.9 \pm 0.7 \arcsec$ for the $C$-band) than at P-band, as well as equal or larger than their respective beam sizes ($\theta_{\rm FWHM} = 5.4 \arcsec \times 5.4  \arcsec$ and $\theta_{\rm FWHM} = 3.4 \arcsec \times 3.0 \arcsec$). Smoothing the $P$-band A-configuration image to the resolution of the VLITE image still shows two peaks with FWHMs similar to the beam size. We note that the larger broadening of the second peak seen in the VLITE data could be due to the fact that we feather B-configuration observations into the A-configuration observations, adding in spatial scales that are not present in the $P$-band A-configuration alone so that it could broaden the Gaussians. The presence of artifacts could also affect the shape of those profiles and influence fitted FWHMs.

Given this new detection of resolved bent jets, our interpretation of the nature of IC 310 must change. We propose the one summarized in the schematic representation of Figure \ref{fig:IC310_schema}. The $\gamma$-ray emission detected from IC 310 must be coming from a region very close to the central supermassive black hole of this galaxy (e.g. \citealt{glawion_black_2016}) and should not follow the bending jets. Therefore, the base of one of the jets, before it starts to bend, must be closely aligned with our line of sight, so that the $\gamma$-rays are beamed toward us.
As shown in Figure \ref{fig:IC310_schema}, we propose to characterize these bent-double radio sources based on different lengths: the tail separation, the thin-jet extent and the radius of curvature of the bent jets ($R_c$). The latter have been commonly used in simulations and observations to describe these sources. From our line of sight, the thin-jet extent can be directly measured from our observations and gives $21 \arcsec = 8$ kpc. However, the radius of curvature obtained from our $P$-band observations is very small, $R_c \sim 5 \arcsec \simeq 2$ kpc. In contrast, the galaxy IC 310 has a reported major and minor axis of $52.1 \arcsec  \simeq 19.5$ kpc and $46.89 \arcsec \simeq 17.6$ kpc respectively (2MASS Extended objects, final release 2003).This direct measurement of the bending radius is, however, likely to be strongly affected by projection effects depending on the unknown angle between the plane of the two jets and our line of sight ($\theta$). Considering this projection effect, the deprojected radius of curvature is given by $R_{c} = R_\mathrm{projected} / \sin(\theta)$ \citep{morsony_simulations_2013}. If this angle is small, the jets will therefore appear projected closer together. Moreover, simulations of AGN jets moving through ICM produce curved jets developing instabilities after 1 bending radius or so (e.g. \citealt{oneill_fresh_2019}). Independently from the $\gamma$-ray detection, this means that IC 310 must be seen in projection and that the viewing angle $\theta$ must be small.
Assuming IC 310 is just like NGC 1265 but seen at a different viewing angle, then comparing the ratios between the thin-jet extent and the tail separation implies that IC 310 must be seen with a $\sim8$ degrees angle between the jet axis and our line of sight. This is consistent with the upper limit on the viewing angle estimated from VLBI observations of the inner jet. For $\theta= 8$ degrees, the deprojected radius of curvature should be around 14 kpc.

According to analytic modeling of bent-double radio sources, the radius of curvature is linked to the properties of the environment as 
\begin{eqnarray}\label{eq}
\dfrac{R_{c}}{h} =  \dfrac{P_\mathrm{jet}}{P_\mathrm{ram}}\, 
\end{eqnarray}
where $h$ is the scale height where the pressure changes, $P_\mathrm{jet}$ and $P_\mathrm{ram}$ are the internal and external pressure, respectively \citep{begelman_twin-jet_1979,jones_hot_1979,burns_dual_1980,odea_constraints_1985}. The scale height $h$ is sometimes interpreted as the jet diameter or the scale height of the interstellar medium of the host galaxy. 
In the case of IC 310, the measured width of the base of the jets is about $5\arcsec \simeq 2$ kpc, whereas the radius of the galaxy is 18 kpc \citep{de_vaucouleurs_third_1991}.
This gives a ratio $R_{c} / h$ of the order of $1-10$, similar to the values obtained from the analytic estimate ($R_{c} / h = 17$) and the simulations ($R_{c} / h$ varies from 0.8 to 14) of bent jets in \cite{morsony_simulations_2013}.
$P_\mathrm{ram}$ can be expressed as the product of the ICM density and the velocity of the galaxy through it: $P_\mathrm{ram} = \rho_\textsc{icm} v^{2}_\mathrm{gal}$. 
In \cite{urban_azimuthally_2014}, the electron density is extracted from \textit{Suzaku} pointings  and is about $n_e \simeq 0.0003 \text{ cm}^{-3}$ at the distance where IC 310 is located ($\simeq 37 \arcmin \simeq 820$ kpc). The ICM density $\rho_\textsc{icm}$ then gives $6 \times 10^{-28}\text{ g/cm}^{3}$, assuming a mean molecular weight of $\mu \simeq 0.61$ and a total number density given by $n_\mathrm{total}\simeq n_e + n_{\text{H}}$, where $n_{\text{H}}$ is the hydrogen density. Then, the radial velocity of IC 310 relative to the surrounding ICM can be calculated from the difference of redshift between NGC 1275 and IC 310, assuming that the ICM is relatively static compared to the motion of NGC 1275. Taking the redshifts measurements from \citep{tully_cosmicflows-3_2016}, $z=0.018850$ and $z=0.017555$ for IC 310 and NGC 1275 respectively, this gives $v_\mathrm{r,gal}= 400 \text{ km/s}$, where IC 310 is moving away from us. The velocity of the galaxy in the plane of the sky is unknown, but it cannot be zero  since the tail of IC 310 is visible and should indicate the direction of infall into the cluster. Therefore, $P_\mathrm{ram} \gtrsim 1 \times 10^{-12}\text{ erg/cm}^{3}$. From equation \ref{eq} and using the ratio $R_{c} / h \sim 1-10$, we therefore estimate the ram pressure into the jet to be  $P_\mathrm{jet} \gtrsim 1- 10 \times 10^{-12}\text{ erg/cm}^{3}$, similar to the pressure of the simulated jets in \cite{morsony_simulations_2013}. For comparison, the internal pressure of the jets in NGC 1265 was also estimated in \cite{odea_astrophysical_1987}, giving $P_\mathrm{jet} \sim 5 \times 10^{-11}\text{ erg/cm}^{3}$.

This interpretation of IC 310 naturally leads to the search for similar cases. There are many head-tail sources showing similar morphology to IC310 \citep{terni_de_gregory_narrow_2017}. Could these sources also be highly projected so that the projected distance between the jets is difficult to resolve?
If this is the case, these could constitute potential sources of $\gamma$-rays emission. Ongoing surveys 
(such as the \textit{Very Large Array Sky Survey}, VLASS, \citealt{lacy_karl_2020},
 the \textit{LOFAR Two-metre Sky Survey}, LoTSS, \citealt{shimwell_lofar_2017} and the \textit{Evolutionary Map of the Universe}, EMU, from the \textit{Australian Square Kilometre Array Pathfinder}, ASKAP, \citealt{norris_emu_2011}) 
will provide deeper, higher-resolution images of numerous head-tail radio galaxies and potentially help resolving other similar cases to IC 310.

Another possibility is that the tail of IC 310 is made of one jet with a transverse structure. There are some examples of such structures in twisted radio jets, such as in the inner part of the Long Tail C in Abell 2256 \citep{owen_wideband_2014}, in the western jet of Cygnus A (e.g. \citealt{perley_jet_1984}) or in M87 \citep{owen_high-resolution_1989}.
Such structures can be created by Kelvin Helmholtz and/or current-driven instabilities \citep{hardee_using_2011}. 

Finally, we look at the X-ray emission in the vicinity of IC 310.  The \textit{Chandra} observations we reprocessed are presented in Figure \ref{fig:image_perseus_ic310_x-ray}. A faint extended X-ray halo with a small elongation towards the tail's direction is visible, and was reported in \cite{dunn_radio_2010}. These minicoronae are common in large elliptical cluster galaxies (\citealt{arakawa_x-ray_2019}. The elongated morphology of the X-ray halo as well as the temperature excess of about $10\%$ in the region between the Perseus cluster center and IC 310 \citep{sato_xmm-newton_2005} both favor the scenario where IC 310 is infalling into the cluster, independently from the radio and gamma-ray observations. This motion would compress the ICM, enhancing the temperature. With the X-ray contours overlaid on the radio emission, we note that the elongation of the minicorona towards the tail seems to be composed of two spurs following approximately the shape of the two bent jets. However, there are other small spurs in the minicorona so it is not clear if there is a real link with the jets.

\begin{figure*}
    \centering
    \includegraphics[height=8cm]{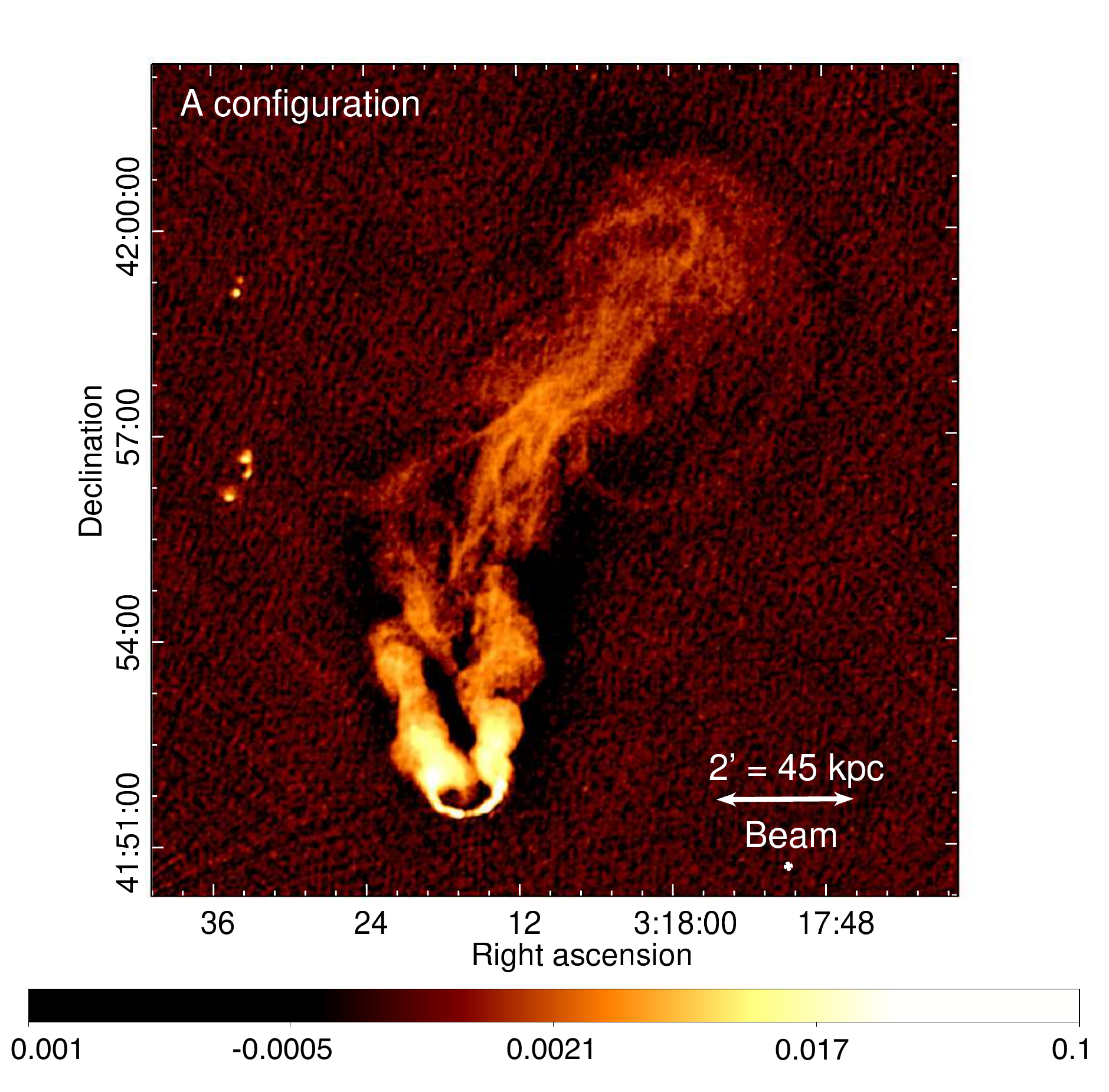}
    \includegraphics[height=8cm]{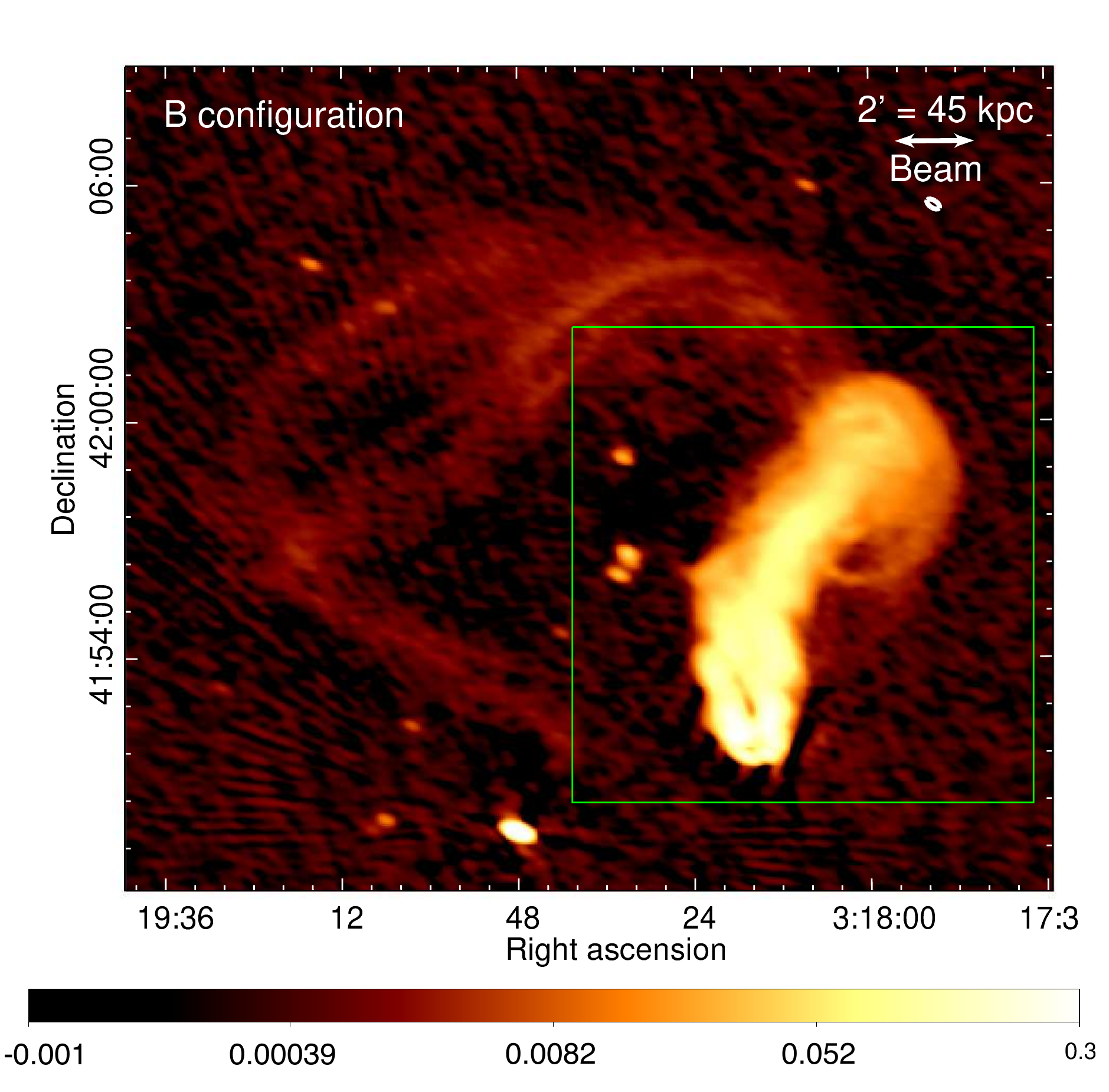}
\caption{The large radio tail of the galaxy NGC 1265 at 230-470 MHz with two different configurations from our VLA observations. Color scales units are Jy beam$^{-1}$. Left - A-configuration observations. Right - B-configuration observations. The green square shows the size of the A-configuration image on the left.}
\label{fig:image_perseus_n1265}
\end{figure*}

\subsection{NGC 1265}\label{NGC 1265_highres}

NGC 1265 is a well-known prominent wide-angle tail radio galaxy in the Perseus cluster. 
Our high-resolution VLA $P$-band image shows two jets emerging from the nucleus that became bent at 90 degrees to the north at a projected distance of $\sim 35 \arcsec \simeq 13$ kpc from their origin (see Figure \ref{fig:image_perseus_n1265} - left). The jets become puffy as the distance from the core increases and merge eventually into a large filamentary tail. Unlike the case of IC 310, the overall shape of the tail is not straight, but bent from the north to the northwest. First discovered by \cite{sijbring_multifrequency_1998}, a much fainter, large extension of the tail is also visible, bending around and to the east by nearly 360 degrees. The whole tail is also visible in our lower resolution B-configuration VLA observations (see Figure \ref{fig:image_perseus_n1265} - right).
This strange morphology is quite puzzling considering the usual interpretation of these tails as tracers of the motion of the galaxy through the ICM. \cite{pfrommer_radio_2011} presented a 3D model based on potential interactions with the cluster gas. The authors noted the strong differences between the bright and the dim parts of the tail in terms of spectral indices (the bright part shows a gradual steepening, while the dim part has a very steep but constant spectral index) and surface brightnesses. They therefore suggested that the radio emission is produced by two separate relativistic electron populations: a more recent population forming the first part of the tail and an older one, having experienced a re-energizing event such as the passage of a shock through it. As in IC 310 and NGC 1272, an asymmetric  minicorona of $\sim 0.6$ keV is also found around NGC 1265, with a sharp edge to the south and an extension to the north, which is interpreted as the result of ram pressure \citep{sun_small_2005}.

The main result from our VLA $P$-band observations of NGC 1265 is the presence of filamentary structures across the entire tail: both bright and dim parts. The A-configuration image reveals a complex network of intricate filaments (see Figure \ref{fig:image_perseus_n1265} - left) while B-configuration observations resolve two pairs of long bending filaments along the dim part of the tail (see Figure \ref{fig:image_perseus_n1265} - right).
The filaments in the bright part were seen before at 1.4 GHz with VLA observations \citep{odea_multifrequency_1986} but not in the extended dim part of the tail detected in the WRST observations from \cite{sijbring_multifrequency_1998} due to their low resolution. 
The width of the smallest filaments, $\sim 1.5$ kpc, is approximately the size of the A-configuration resolution, so they could be even thinner. 
It seems that the emission mechanism at play is similar for both parts of the tail and that it gives rise to filamentary structures.  

The filamentary structures detected in NGC 1265 are very similar to what is found in 3C 129, another prototype tailed radio galaxy (e.g. \citealt{lane_3c_2002}). As mentioned earlier, a complex web of thin filaments is also seen inside the radio lobes of M87 \citep{owen_m87_2000}. Similar filaments are also resolved in the ``Large Relic'' in Abell 2256, a Mpc-scale diffuse radio structure north of the cluster center \citep{clarke_deep_2006,owen_wideband_2014}. Moreover, several radial filaments are seen in the mini-halo of Perseus from our B-configuration 270-430 MHz radio map \citep{gendron-marsolais_deep_2017}. In contrast, such filaments seem to be absent from most halos, another class of diffuse cluster radio sources found only in highly-perturbed clusters and extending to Mpc scales. This might indicate a fundamental difference between these sources. However, this might also be an observational limitation as giant halos have much lower surface brightness than mini-halos. For example, the halo found in MACS J0717.5+3745 shows evidence of filamentary substructures with sizes of 100-300 kpc in deep VLA observations \citep{van_weeren_chandra_2017}. As suggested in \cite{gendron-marsolais_deep_2017}, the filaments could trace regions of enhanced magnetic fields or turbulence, or reflect the original distribution of fossil plasma left by an old AGN outburst, that is being re-accelerated by turbulence or weak shocks. However, without available data at other frequencies with similar depth and resolution to our observations - enabling the whole tail of NGC 1265 to be visible - no spectral analysis can yet be done on the faint part of the tail and it remains difficult to explain the nature of this structure.

\begin{figure*}
\centering
\includegraphics[height=8cm]{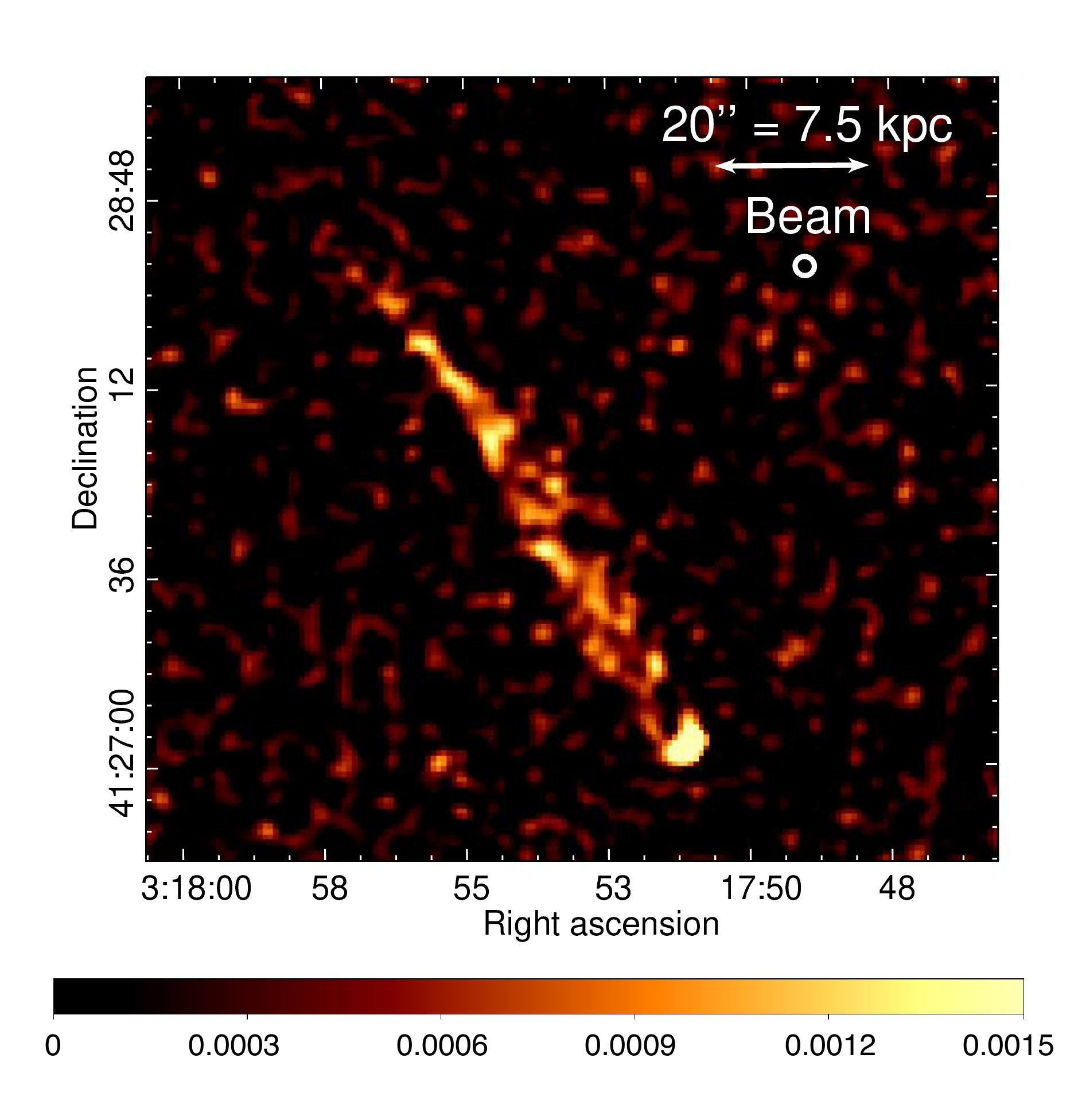}
\includegraphics[height=8cm]{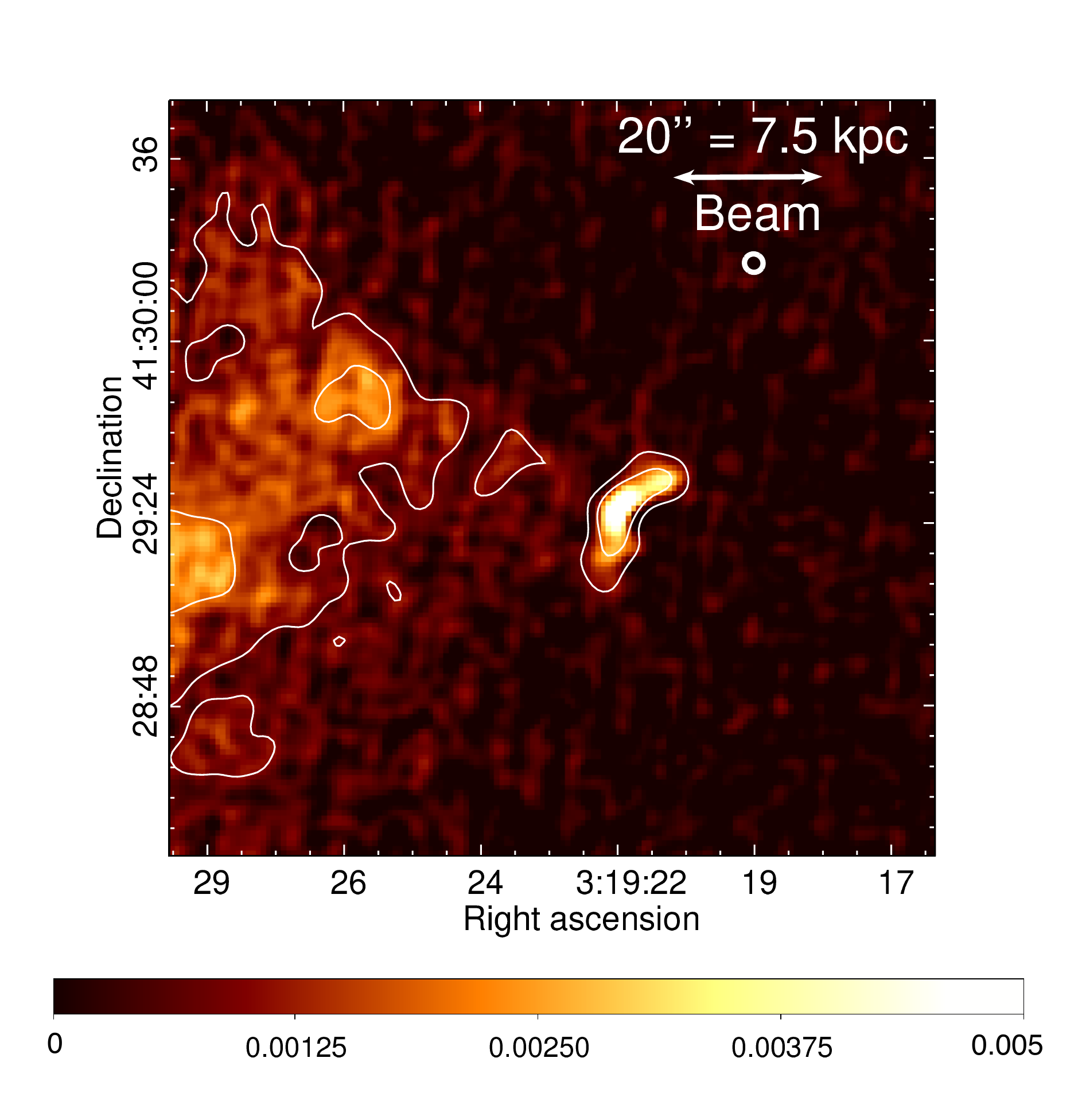}
\caption{Color scales units are Jy beam$^{-1}$. Left: A-configuration 230-470 MHz VLA observations of CR 15.
Right: A-configuration VLA 230-470 MHz observations of NGC 1272. Logarithmic contours start from $3\sigma = 0.81$ mJy beam$^{-1}$ to 1 Jy beam$^{-1}$ (10 contours levels are shown). The faint emission seen on the left is thought to be part of the mini-halo emission.}
\label{fig:image_perseus_cr15_n1272}
\end{figure*}

\subsection{CR 15}\label{CR15_highres}

The small head-tail source CR 15 is located between NGC 1275 and IC 310, its tail pointing in a northeast direction, away from the center of the Perseus cluster. The host galaxy associated with the radio source ($z=0.015$, \citealt{miller_radio_2001}) was first identified as a Perseus cluster member by \cite{chincarini_dynamics_1971}, where its name (CR 15) comes from. \cite{miley_active_1972} used the direction of this tail to prove that the other head-tail radio sources detected in the Perseus cluster were not created by an intergalactic wind coming from NGC 1275. 
In our observations, the faint tail extends up to $1.7 \arcmin \simeq 38.25$ kpc from the core and is very narrow and straight (see Figure \ref{fig:image_perseus_cr15_n1272} - left). Fitting a Gaussian profile across the width of the brightest part of the tail reveals a FWHM of $10 \arcsec \simeq 3.8$ kpc, so our observations do resolve the tail width. The tail gets thinner as the distance from the core increases, until its width is about the size of the beam. Compared to other head-tail sources, the tail of CR 15 is quite narrow, but is also not the narrowest known tail. For example, the ``Long Tail C'' in Abell 2256 is even thinner, $\lesssim 100$ pc near the core \citep{owen_wideband_2014}. 
The morphology of CR 15 is similar to IC 310 but this time our observations show no indication of a double jet structure. CR 15's tail is of similar extent but about 10 times fainter in terms of surface brightness and somewhat narrower than the brightest part of the tail of IC310 (IC 310 merged tail has a width of $\sim 25 \arcsec \simeq 10$ kpc). As there is no $\gamma$-ray detection from CR 15, this means either that we are not looking down towards the core as we do for IC 310, or that there are much less high energy emission generated in its AGN. If the former is true, then both interpretations of CR 15 as either a blazar-like single Doppler boosted jet or an IC 310-like source (as described in Figure \ref{fig:IC310_schema}) are incorrect and the nature of CR 15 is unclear. Moreover, interpreting these straight, narrow, one-sided radio tails (such as CR 15 and Abell 2256 tail C) as head-tails (merged bent jets close to the plane of the sky) is problematic since it implies that the jets stay collimated on tens of kpc after being bent, which is not what is typically observed in wide or narrow-angle tail sources, nor in simulations \citep{morsony_simulations_2013,oneill_fresh_2019}.

\begin{figure*}
\centering
\includegraphics[width=7cm]{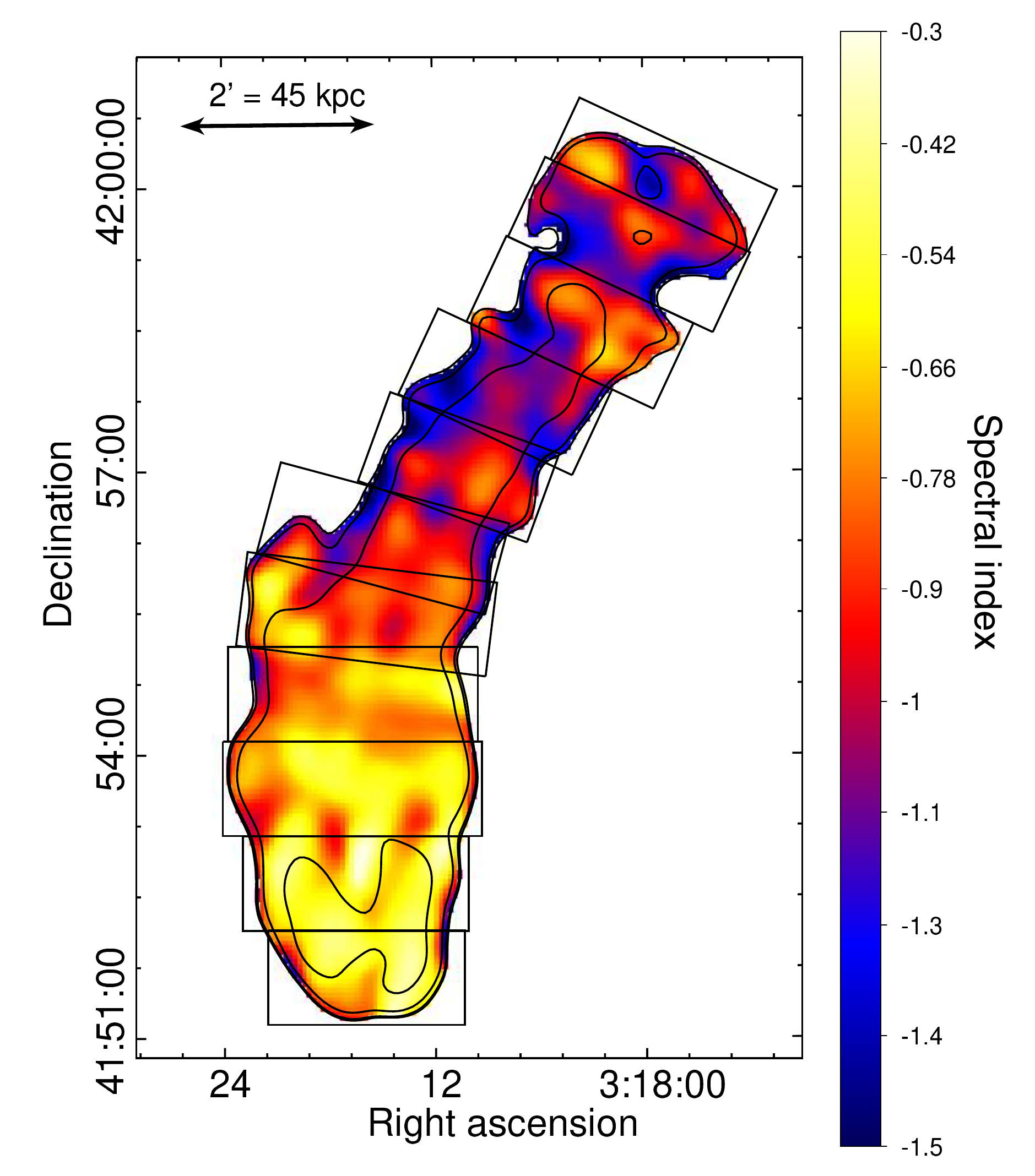}
\includegraphics[width=7cm]{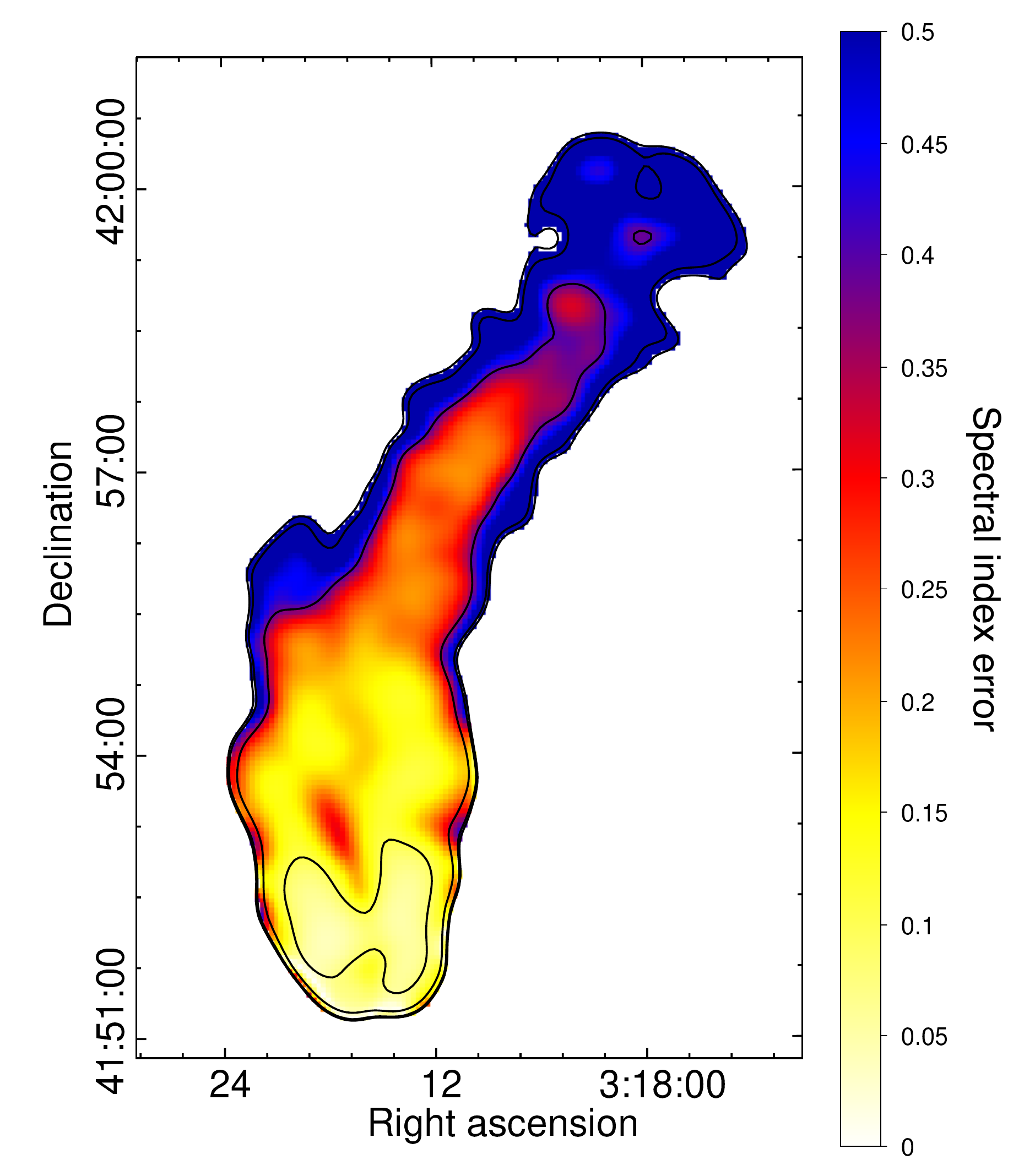}
\includegraphics[width=7cm]{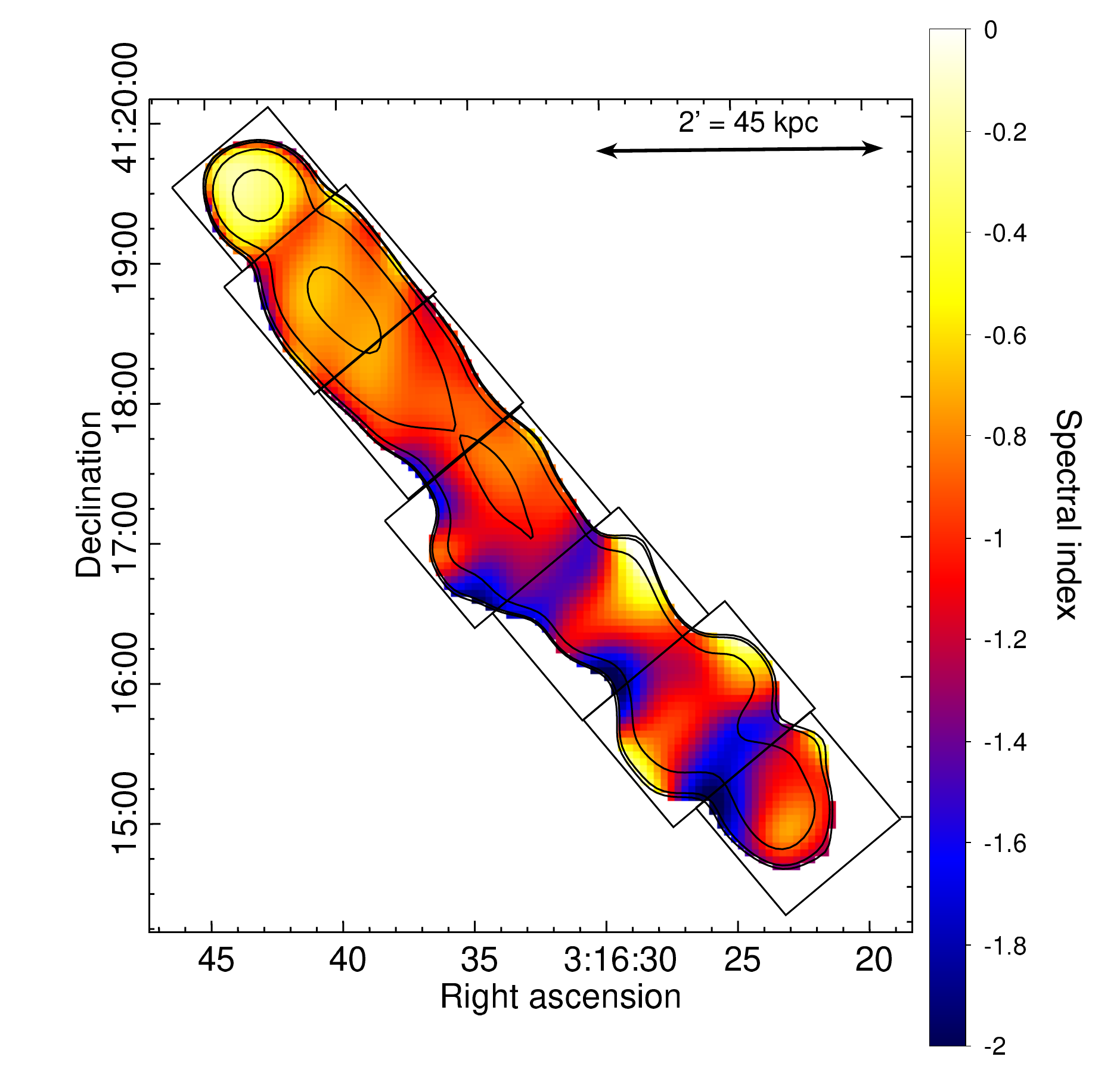}
\includegraphics[width=7cm]{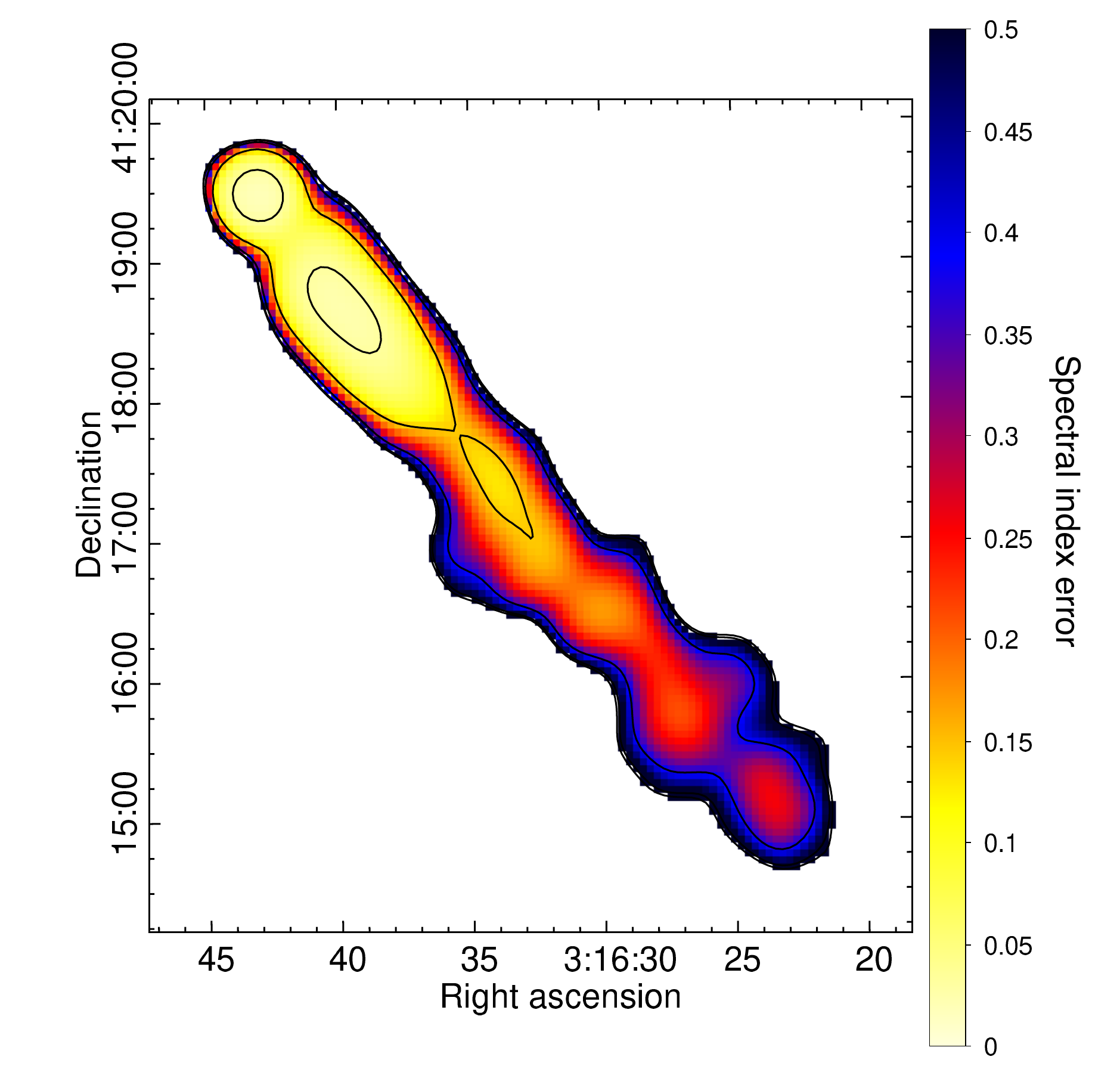}
\caption{The spectral index maps (left) and their corresponding error maps (right) of the head-tail sources NGC 1265 and IC 310 produced with our B-configuration 230-470 MHz VLA observations combined with the GMRT 610 MHz observations of these sources presented in \protect\cite{sebastian_giant_2017}. Logarithmic contours of the GMRT images used are overlaid on each maps, starting from $3\sigma$ to 1 Jy beam$^{-1}$ for NGC 1265 and up to 0.1 Jy beam$^{-1}$ for IC 310 (5 contours levels are shown). The profiles extracted from the regions shown on the spectral index maps are presented in Figure \ref{fig:image_alpha}.}
\label{fig:alpha_maps}
\end{figure*}

\subsection{NGC 1272}\label{NGC 1272_highres}

The discovery of bent double radio jets in NGC 1272 was first reported in \cite{mcbride_bent_2014} using archival VLA observations at 1.4, 1.8 and 3.2 GHz. Located in projection at only $\sim 5\arcmin \simeq 110$ kpc from the nucleus in NGC 1275, it is positioned at the western edge of the mini-halo emission. Being at a closer projected distance from the cluster center than most of the other bent radio galaxies and moving at least at 1450 $\text{km s}^{-1}$ relative to the surrounding ICM, it must undergo strong ram pressure stripping \citep{mcbride_bent_2014,arakawa_x-ray_2019}.  
As with IC 310 and NGC 1265, NGC 1272 is also surrounded by a cool soft X-ray corona of 0.63 keV temperature and 1.2 kpc radius \citep{arakawa_x-ray_2019}. The reason why these minicoronae are not being stripped away by the ICM is not well understood. However, simulations indicate that a cool, $< 1$ keV component can remain, probably due to its lower entropy, the high density contrast with the ICM and a large standoff distance from the shock (e.g. \citealt{sheardown_recent_2018,zhang_standoff_2019}). The presence of magnetic fields could also play a role in the suppression of the depletion of the minicoronal gas (e.g. \citealt{dursi_draping_2008,arakawa_x-ray_2019}).

With our high-resolution VLA observations, we also report the presence of bent double jets at 230-470 MHz (see Figure \ref{fig:image_perseus_cr15_n1272} - right). Both jets show similar brightness. The northern one curves smoothly toward the west while the southern one takes a bend to the east. \cite{mcbride_bent_2014} estimated the radius of curvature of the jets in NGC 1272 to be $R \sim 2$ kpc by fitting visually a circle to the shape of the double jets. 
As this estimate falls well within the half-light radius of the galaxy ($\sim 10$ kpc), this means either that the influence of the ICM acts well within the galaxy or that its jets are closely aligned with the plane of the sky.
Moreover, NGC 1272 is moving toward us at a 1450 $\text{km s}^{-1}$ radial velocity relative to the ICM \citep{arakawa_x-ray_2019}. If we interpret the bending of the jet as pointing toward the direction of its current orbital motion, then it means that NGC 1272 is getting out from behind the cluster, traveling through the cluster towards the observer, from the west to the east. Because of this perspective, the tail might appear foreshortened. As discussed earlier, the link between NGC 1272's jets and the nearby mini-halo is not clear from our observations.

\begin{figure}
\centering
\includegraphics[width=\columnwidth]{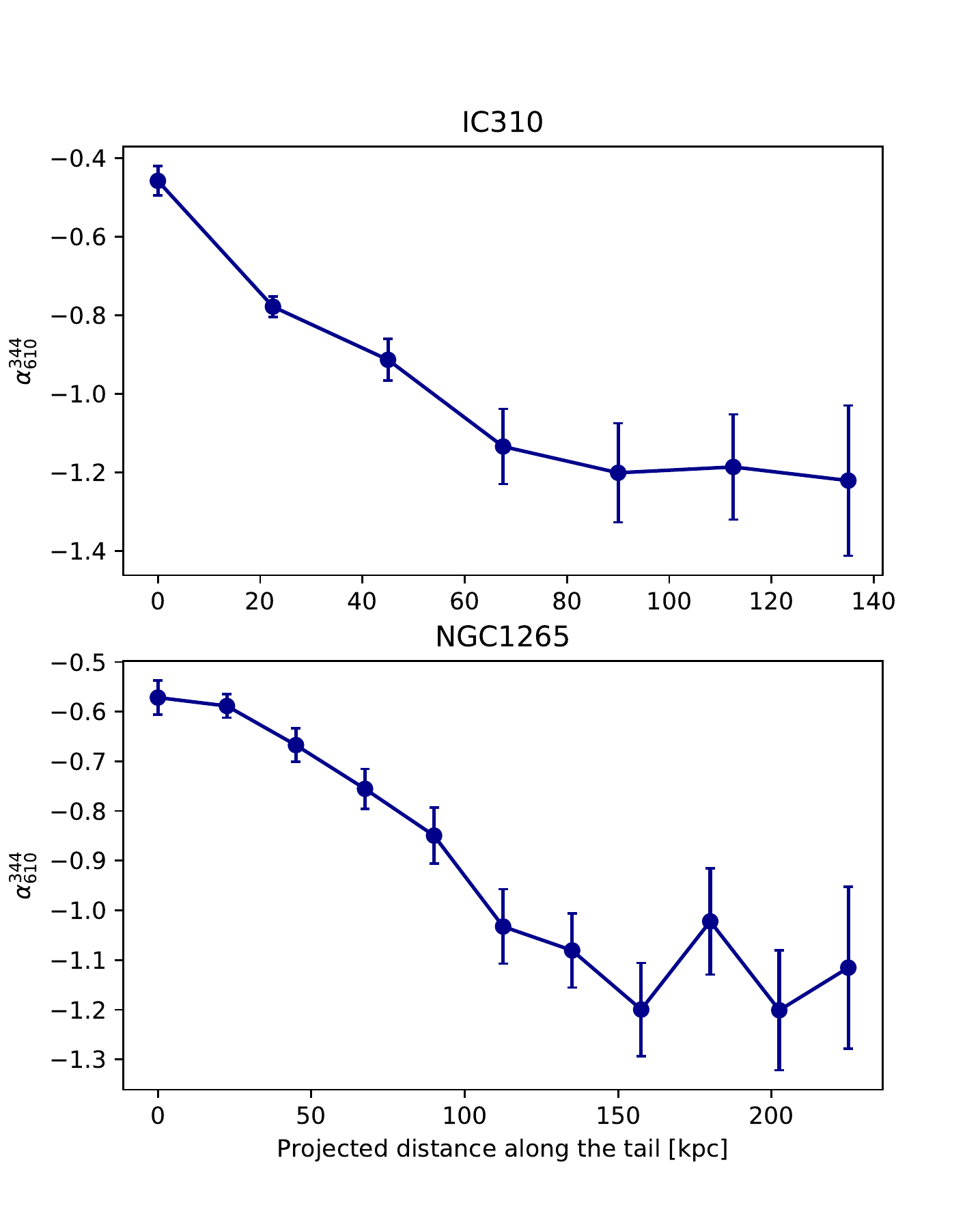}
\caption{The spectral index profile across the head-tail sources NGC 1265 and IC 310 based on the regions showed in Figure \ref{fig:alpha_maps} produced with our B-configuration VLA observations (with effective frequency of 344 MHz) combined with the GMRT 610 MHz observations of these sources presented in \protect\cite{sebastian_giant_2017}. In addition to the statistical errors plotted, there is a potential for up to $\sim 10\%$ systematic error, due to flux scale errors, which would result in a uniform shift up or down of all spectral indices by $\sim0.25$.}
\label{fig:image_alpha}
\end{figure}

\section{Spectral index maps of NGC 1265 and IC 310}\label{Spectral index maps of NGC 1265 and IC 310}

We now discuss the radio spectral characteristics of NGC 1265 and IC 310. 
In the presence of a magnetic field, a population of relativistic electrons with energy distribution following a power law $N(E) \propto E^{-p}$, where $N$ is the number of electrons with energy $E$, will emit synchrotron radiation following a power law emission spectrum $S_{\nu} \propto \nu^\alpha$. The spectral index $\alpha$ is linked to the index of the energy distribution $p$ as $p = 1-2\alpha$. With time, the electrons will lose energy through this synchrotron emission as well as through inverse Compton losses. The losses from both mechanisms are proportional to the square of the energy of the electrons, the highest energy electrons are therefore affected first (e.g. \citealt{sijbring_multifrequency_1998}). This evolution in the energy distribution will affect the spectrum's shape.

In order to study these spectral characteristics, we produced spectral index maps combining our B-configuration VLA observations (with effective frequency of 344 MHz) with the GMRT observations at an effective frequency of 610 MHz of NGC 1265 and IC 310 presented in \cite{sebastian_giant_2017}.
These maps are presented in Figure \ref{fig:alpha_maps}. Uncertainties on the spectral indices are calculated using standard propagation of errors. We also extracted profiles across both head-tail sources (see the regions defined in Figure \ref{fig:alpha_maps} and profiles in Figure \ref{fig:image_alpha}), following the curvature of the tail in the case of NGC 1265 and extracting individual fluxes from the two input images in each region. In addition to the statistical errors plotted, there is a potential for up to $\sim 10\%$ systematic error, due to flux scale errors, which would result in a uniform shift up or down of all spectral indices by $\sim0.25$. Overall, for both sources, the spectral index becomes steeper as the distance from the head increases. Variations from this decrease are small compared to the error bars and we do not think they are significant. Variations are also seen in the maps, towards the fainter part of both tails but are associated with residual artifacts present in the GMRT images which increase the uncertainties as shown on the spectral index error maps.
In general, these maps and profiles are consistent with the previous ones obtained in \cite{feretti_electron_1998,sijbring_multifrequency_1998,sebastian_giant_2017} at similar frequencies. Since we have images at only two frequency bands (344 MHz and 610 MHz), it is hard to constraint the evolutionary models describing the spectral falloff at high frequencies.
Current injections models such as the Jaffe-Perola model (JP, \citealt{jaffe_origin_1977}), the Kardashev-Pacholczyk model (KP, \citealt{kardashev_nonstationarity_1962,pacholczyk_radio_1970}), the continuous injection model (CI, \citealt{pacholczyk_radio_1970}) or the Komissarov-Gubanov model (KGJP/KGKP, \citealt{komissarov_relic_1994}), cannot be applied but such analysis will be undertaken in a future paper with additional images at other radio frequencies having similar resolution to our data.

\section{Conclusion}\label{Conclusion_highres}

We presented a high-resolution radio map of the Perseus cluster obtained from 5 hours of observations with the VLA at 230-470 MHz in the A-configuration. The combination of high dynamic range ($27 000:1$) and resolution (beam size of $\theta_{\rm FWHM} = 3.7 \arcsec \times 3.6 \arcsec$) achieved has allowed the identification of new structures, both in the central regions surrounding NGC 1275 and in several complex radio sources harbored in the Perseus cluster. Our main conclusions are:
\begin{enumerate}
\item New hints of sub-structures are seen in the inner radio lobes, showing a patchy/filamentary nature, rather than uniform emission. The faint spurs of emission extending into both of the outer X-ray cavities are seen for the first time at these frequencies and are consistent with spectral aging. Our high-resolution observations also resolve the northwestern spur into a double filamentary structure. 
\item Only some of the brightest western parts of the extended mini-halo emission remain visible at this high resolution, near NGC 1272 and its bent double jets. Deeper, higher-resolution imaging with spectral studies of this emission are needed to attest the potential connection between NGC 1272 and the mini-halo emission and its possible role in the creation of such structure.
\item We report the presence of filamentary structures across the entire tail of NGC 1265: the already known network of filaments in its brightest part, but also two pairs of long filaments in the faintest bent extension of the tail. Some of these filaments are unresolved ($\sim 1.5$ kpc wide, corresponding to our resolution). Despite the differences in terms of surface brightnesses, it seems that the emission mechanism at play here is similar for both parts of the tail and that it gives rise to filamentary structures. Such filaments have also been seen in other cluster radio sources such as relics and radio lobes, indicating that there may be a fundamental connection between these radio structures.
\item For the first time, our observations reveal the tails of IC 310 to be two distinct, narrowly-collimated jets, consistent with the original interpretation of a narrow-angle tail radio galaxy infalling into the cluster. Both the small radius of curvature of its bent jets and the detection of $\gamma$-rays from its nucleus are consistent with a narrow-angle tail galaxy seen at an angle of about 10 degrees between the plane of the bent jets and our line of sight. With their high spatial resolution and low level of noise, our observations therefore provide important constraints on the nature of this source and imply that blazars and bent-jet radio galaxies are not mutually exclusive.  
\item We resolve the very narrow and straight tail of CR 15, extending up to $1.7 \arcmin \simeq 38.25$ kpc from its core, but do not see any indication of a double jet structure, so that the interpretation of the nature of such head-tail sources is yet unclear. 
\end{enumerate}

\section*{Acknowledgments}
JHL is supported by NSERC through the discovery grant and Canada Research Chair programs, as well as FRQNT. RJvW acknowledges support from the ERC Starting Grant ClusterWeb 804208. Partial support for LR comes from U.S. National Science Foundation grant AST 17-14205 to the University of Minnesota. Basic research in radio astronomy at the Naval Research Laboratory is supported by 6.1 Base funding. The National Radio Astronomy Observatory is a facility of the National Science Foundation operated under cooperative agreement by Associated Universities, Inc. 

\section*{Data availability}
The calibrated data generated in this article will be shared on reasonable request to the corresponding author.



\bibliographystyle{mnras}
\bibliography{biblio_perseus}




 

\bsp	
\label{lastpage}
\end{document}